\documentclass[12pt, draftcls, onecolumn]{IEEEtran}
\usepackage{epsfig,graphicx}
\usepackage{nomencl}
\makenomenclature
\usepackage{array}
\usepackage{lscape}
\usepackage{epsfig,graphicx}
\usepackage{float}
\usepackage{color}
\usepackage{algorithm}
\usepackage{algorithmic}
\usepackage{amssymb}
\usepackage{subfigure}
\usepackage{multirow}
\usepackage{colortbl}
\usepackage{slashbox}
\ifCLASSINFOpdf
\else
  \graphicspath{{../eps/}}
\fi
\begin{document}

%
\title{UROP: A Simple, Near-Optimal Scheduling Policy for Energy Harvesting Sensors}
%
%
%

\author{Omer~Melih~Gul,
         and~Elif~Uysal-Biyikoglu,~\IEEEmembership{Senior Member,~IEEE}
\thanks{O. Melih Gul and Elif Uysal-Biyikoglu are with the Department of Electrical and Electronics Engineering, Middle East Technical University, Ankara, Turkey (omgul@metu.edu.tr, uelif@metu.edu.tr). A preliminary version of this paper has been accepted to appear at the 2014 IEEE Wireless Communications and Networking Conference (WCNC 2014), Istanbul, Turkey.}
\thanks{}
\thanks{}}

%
%

\markboth{IEEE TRANSACTIONS ON INFORMATION THEORY (SUBMITTED PAPER)}%
{Shell \MakeLowercase{\textit{et al.}}: Bare Demo of IEEEtran.cls for Journals}
%



\maketitle

\begin{abstract}

This paper considers a single-hop wireless network where a central node (or fusion center, FC)
collects data from a set of $m$ energy harvesting (EH) nodes (e.g. nodes of a wireless sensor network). In each time slot, $k$ of $m$ nodes can be scheduled by the FC for transmission over $k$ orthogonal channels. FC has no knowledge about EH processes and current battery states of nodes; however, it knows outcomes of previous transmission attempts. The objective is to find a low complexity scheduling policy that maximizes total throughput of the data backlogged system using the harvested energy, for all types (uniform, non-uniform, independent, correlated (i.e. Markovian), etc.) EH processes. Energy is assumed to be stored losslessly in the nodes’ batteries, up to a storage capacity (the infinite capacity case is also considered.) The problem is treated in finite and infinite problem horizons. A low-complexity policy, UROP (\emph{Uniformizing Random Ordered Policy}) is proposed, whose near optimality is shown. Numerical examples indicate that under a reasonable-sized battery capacity, UROP uses the arriving energy with almost perfect efficiency. As the problem is a restless multi-armed bandit (RMAB) problem with an average reward criterion, UROP may have a wider application area than communication networks.

\end{abstract}


%
\IEEEpeerreviewmaketitle

\section{Introduction}
%
%
%
%
\IEEEPARstart{P}{OWER} resource and battery lifetime are important issues for networks such as Wireless Sensor Networks (WSNs). Energy harvesting (EH) \cite{1} can enable WSN operation in environments where maintenance is impractical or too costly. Energy harvesting (EH) extends reliable operation lifetime \cite{2}, \cite{3}. Energy may be harvested from the environment in many different ways (solar, kinetic, etc.) \cite{4}. Since energy harvesters generally depend on uncontrollable energy resources and the amount of harvested energy is generally low \cite{4}, WSNs need robust, environmentally adaptive, energy efficient policies for their operations. 

In this paper, we consider a WSN where a fusion center (FC) collects data from $m$ EH sensor nodes by assigning the nodes to $k$ orthogonal communication channels in each time slot. It is assumed that each node always has data to transmit (i.e., nodes are data backlogged). Each node has a battery (of a certain capacity, and without leakage) to store harvested energy. It is also assumed that the multi-access communication is error-free and there is no fading. If a node is scheduled, it will be assigned one of the channels. When a node is scheduled to transmit, it can transmit data to the FC if it has sufficient energy to send a packet. The transmission of each packet lasts an entire time slot. The objective of the FC is to maximize the total throughput over a finite or infinite problem horizon.

In practice, battery states of nodes could be made available to the FC through some additional cost (i.e. feedback) and complexity. However, it is interesting from a practical perspective to consider the case where the FC makes scheduling decisions without knowledge of the instantaneous battery states at nodes, or their statistics. Fortunately, it turns out that this lack of knowledge has little effect on performance. We will observe that by knowing only the outcomes of previous transmission attempts, the FC can schedule almost as efficiently as an omniscient scheduler.

This problem may be formulated as a partially observable Markov Decision Process (POMDP), and dynamic Programming (DP) \cite{5} can be employed for optimal solution. However, DP has exponential complexity with respect to number of nodes $m$ \cite{5}. Furthermore, the state space of DP should be very large to get a good approximation to the problems with continuous state variables like energy. Therefore, complexity of DP becomes excessively high for the EH scheduling problem with large number of nodes.

A second approach for solving this scheduling problem is reinforcement learning by considering the problem as a POMDP.  Q-learning \cite{6} is the easiest to implement and the most effective model-free algorithm among reinforcement learning algorithms. Q-learning guarantees convergence to optimal for a generic model. However, Q-learning is not applicable for problems with large-state space because its convergence is slow \cite{7}. In fact, many algorithms can guarantee the convergence to optimal behavior \cite{8}. However, in many practical applications, a policy which achieves near optimality quickly is preferable to the policy which converges slowly to exact optimality \cite{7}. As the discount factor gets closer to 1 (i.e. the undiscounted case), the convergence rate of Q-learning decreases more. There are approaches such as R-learning \cite{9} which maximize average reward; however, the convergence of R-learning has not been proven. Also, reinforcement learning has a very important problem: \emph{the trade-off between exploration and exploitation} \cite{10}. Therefore, Q-learning and generally reinforcement learning do not seem to be suitable for obtaining an efficient and practical solution to this scheduling problem, especially a large number of sensors and a continuous state variable, energy, is considered.

Another approach for this scheduling problem is to consider it as a restless multi-armed bandit problem (RMAB) which is a special version of POMDP. RMAB is an extension to classical multi-armed bandit problem which is solved optimally by Gittins \cite{11} and an optimal solution is proposed under certain assumptions by Whittle \cite{12}. Papadimitriou and Tsitsiklis show that finding optimal solution to a general RMAB is PSPACE-hard and it has a very high computational complexity \cite{13}. Considering memory limits of sensors, a much more applicable policy is required. Therefore, a simpler approach called a myopic policy (MP) is suggested for RMAB problems and proven to be optimal in limited cases for the sensor management problems in \cite{14}, \cite{15}, \cite{16}. However, a myopic policy is not generally optimal since MP concentrates only on the present and not consider the future \cite{17}, \cite{18}. A channel probing problem is studied in \cite{19} and it is shown that MP is not always optimal. The assumption that the scheduling decision does not affect transition probabilities  was an appropriate one for the problems addressed in \cite{14,15,16,19}. However, for the EH scheduling problem at hand, this is not a reasonable assumption, as energy is a flexible resource that can be stored without any discount (ignoring battery leakage which is very minor in practice \cite{2}) and can be used whenever desired.  Therefore, the solutions presented in \cite{14,15,16,19}  papers are not directly applicable to our problem.

The closest works in the literature to the problem at hand are the scheduling problems studied in \cite{18,20}. We have posed essentially the same problem, with the exception that no battery and unit sized batteries at nodes are assumed in \cite{18} and \cite{20},respectively. In both \cite{18} and \cite{20}, the scheduling problem is formulated as a POMDP where the focus is on immediate reward instead of future rewards. In \cite{20}, a single-hop wireless sensor network which consists of EH transmitter nodes with a unit sized battery and a central receiver node with multi server is considered as a restless multi-armed bandit problem (RMAB). Optimality of Whittle index policy which is generally suboptimal for RMAB \cite{21} is proven for a certain case under certain assumptions on the EH process. In \cite{20}, the optimality of a Round-Robin based myopic policy is proved under the assumption that each node has only unit sized battery and the ratio between the number of transmitter nodes and the number of communication channels of the central node is an integer ($m/k$ is an integer). In \cite{18}, the problem is formulated as POMDP and the optimality of MP is proven for two cases: 1) the nodes are not able to harvest and transmit simultaneously, and the EH process transition probabilities are affected by the scheduling decisions, and 2) the nodes have no battery. Since myopic policies proposed in \cite{18} and \cite{20} are based on Round-Robin(RR) Scheme, assuming that $p=m/k$ is an integer is important (also period of RR Policy).These assumptions are somewhat restrictive for real life implementation.

To set up the problem, a model about the generation and usage of energy is needed. First, energy in a node's battery decreases if the node sends a data packet. Second, energy in a battery increases in a continuous fashion by harvested energy. Third, battery leakage is neglected. This assumption follows from examining typical batteries in use today for which leakage is negligibly small for over durations of several minutes. Based on these mild assumptions about energy, a suitable performance measure for a policy can be average reward over the finite and the infinite horizon rather than expected total  discounted reward for this scheduling problem which is a delay-insensitive communication problem \cite{22}. In communication network problems, delay issue is investigated as average delay not as discount. In applications, EH sources may use vibrational or kinetic energy, the behavior of which is typically not predictable \cite{1}, \cite{3}. Optimal scheduling for this continuous, independent EH process becomes a hard problem, and the problem requires good near-optimal solutions. 

By taking a deterministic approach, a near-optimal transmission scheduling policy, Uniforming Random Ordered Policy (UROP), is developed by assuming that each sensor has an infinite capacity battery (It will be shown that if the sensors have a reasonable-sized finite battery, UROP has almost same efficiency as its efficiency under a reasonable-sized finite battery assumption). It is also guaranteed that UROP is \emph{asymptotically optimal} for a general case of energy arrival process under the infinite battery assumption (larger than unit battery) as the horizon length increases. In comparison with the myopic policies in \cite{18} and \cite{20}, UROP can still guarantee near-optimal performance when $p=m/k$ is not an integer.

The rest of this paper is organized as follows. The system model and problem formulation are described in Section II. In Section III, we study the scheduling capacity. In section IV, we show that Round-Robin based policies cannot guarantee 100\% throughput under many non-uniform energy harvesting process for nodes. We show the optimal omniscient solution for this problem in section V. In section VI, we suggest a novel, low-complexity scheduling policy which is nearly throughput optimal for quite general EH processes (uniform, non-uniform, independent, correlated) in a finite horizon problem under an infinite battery assumption. Next, efficiency bounds on UROP are obtained. Section VII extends the results from finite horizon to infinite horizon. In section VIII, we compare the performance of UROP with that of a Round-Robin policy and the Myopic Policy in \cite{18,20} through simulations. Section IX concludes the paper.


\section{System Model and Problem Formulation}

We consider a single-hop wireless network in which $m$ energy harvesting (EH)-capable sensors have circularly symmetric distribution around a Fusion Center (FC) and send data packets to FC 
(see Figure 1). The WSN operates in a time-slotted fashion over time slots (TSs) of equal duration. In each TS, FC schedules $k$ of $m$ sensors for data transmission by assigning each one of $k$ orthogonal channels. We assume that each sensor always has data to transmit (i.e. data is backlogged as in \cite{18} and \cite{20}) during the problem horizon of $N$ TSs. Data packets have equal size and require \emph{unit energy} for transmission.  
\begin{figure}
	\centering	\includegraphics[width=0.40\textwidth]{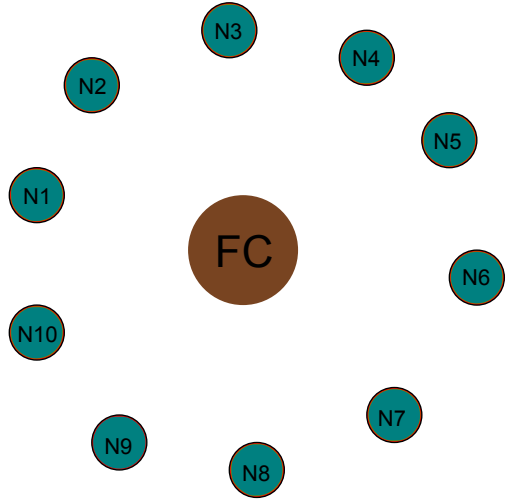}
	\caption{A single hop wireless sensor network where a fusion center (FC) collects data from energy harvesting (EH) nodes located in a star topology around it.}
	\label{Figure 1}
\end{figure}

The EH processes are assumed to be independent for each node. The total energy harvested by sensor $s_i$ by time $t$ is denoted by $E_i^{tot}(t)$, and the increment of this energy harvested during the TS $t$ is denoted as $E_i^h(t)$. The energy present in the battery at $t$ (stored minus used) is $B_i (t)$.  

We denote by $S_m=\left\lbrace s_1,s_2,..,s_m\right\rbrace$ and $A_k=\left\lbrace a_1,a_2,..,a_k\right\rbrace$, the set of all sensors and the set of orthogonal channels, respectively. The amount of data sent by node $s_i$ in TS $t$ is denoted by $D_i (t)=1(s_i\in{S(t)})1(B_i(t)\geq1)\in{\left\lbrace 0,1\right\rbrace}$ where $1(A)$ is the indicator function of event A, and $S(t)\subset{S_m}$ is the set of $k$ nodes scheduled at $t$. The set $S(t)$ is determined by a policy $\pi$. 

Two definitions are in order: A \emph{fully efficient policy} (alternatively, a \emph{$100\%$ efficient policy}) ensures that the nodes use up all of the harvested energy by the end of the problem horizon, more precisely, $B_i(N)<1$ for $\forall s_i \in S_m$. An \emph{optimal policy} is one that maximizes data throughput for the given sequence of energy harvests. For certain energy harvest processes, an optimal policy may not be fully efficient, as we will be clear in the next section.


Consistently with previous literature, the general objective is to maximize the expected discounted reward over the problem horizon:
\begin{equation}\label{1}
\max_{S(t), t=1,.., N} V_{\rm{tot}}(t)=\max_{s(t), t=1,.., N}\rm{E}[\sum_{t=1}^{N}\beta^{t-1}\sum_{s_i\in S_m}1(s_i\in S(t)) 1(B_i(t)\geq1)]
\end{equation} 
where $0<\beta\leq1$ is the discount factor, which reduces the value of data sent later. 
The discount factor corresponds to placing lower value on data that is delayed. However, note that the problem at hand assumes infinite backlog and is therefore delay insensitive by nature. 
The discount could also be considered to model battery leakage that happens as transmission is withheld. Therefore, average reward criterion is more suitable measure for delay-insensitive communication problems like this scheduling problem than discounted reward criterion \cite{22}.  
Consistently with our assumptions about infinite data buffers, and no battery leakage, we shall set $\beta=1$ and convert the objective function in (\ref{1}) to that in (\ref{2}), which is an average reward criterion.
\begin{equation}\label{2}
\max_{S(t), t=1,.., N}\frac{V_{\rm{tot}}(t)}{N}=\max_{S(t), t=1,.., N}\rm{E}[\frac{1}{N}\sum_{t=1}^{N}\sum_{s_i\in S_m}1(s_i\in S(t)) 1(B_i(t)\geq1)]
\end{equation}
We propose an algorithm, UROP, which achieves nearly 100\% throughput (and 100\% efficiency whenever a fully efficient schedule is feasible) in a broad class of energy harvesting (arrival) process under infinite battery assumption. In this work, efficiency of a policy $\pi^*$ ($\eta_*$) is defined as ratio of total throughput by the policy $\pi^*$ to total throughput by fully efficient policy $\pi^{fe}$ on the problem horizon ( $\eta_{fe}=1$). In section III, it is proven that efficiency of an arbitrary Round Robin Policy with quantum=1 $\pi^{RR}$ is very close to that of myopic policy $\pi^{MP}$ ($\eta_{MP}$) proposed in \cite{18} and \cite{20}. Therefore, the efficiency of UROP $\pi^{UROP}$ ($\eta_{UROP}$) will be compared with that of an arbitrary Round Robin Policy with quantum=1 $\pi^{RR1}$ ($\eta_{RR}$) in section VII for simplicity. 

An arrival process is called \emph{admissible} if a fully efficient schedule is possible. By the analogy with admissible processes in these problems, we introduce four new terms which we use for the EH scheduling problem in the rest of paper. \emph{Density of sensor i}, $D_i$, is the total number of packets sent by the sensor $s_i$ with $\pi^{fe}$ normalized by $\frac{kN}{m}$ during problem horizon $N$. \emph{Partial Density of sensor} $s_i$, $D_i^{(T)}$, is the total number of packets sent by the sensor $s_i$ with $\pi^{fe}$ normalized by $\frac{k(N-T)}{m}$ in the interval $[T,N]$. \emph{Density} $(D)$ is the average of densities of all sensors during problem horizon $N$ (see Equation (\ref{3})). \emph{Partial Density} ($D^{(T)}$) is the average of partial densities of all sensors in the interval $[T,N]$ (see Equation (\ref{4})). By definition, $D,D^{(T)}\leq 1$. 
\begin{equation}\label{3}
D=\frac{\sum_{s_i\in S_m}D_i} {m}
\end{equation}
\begin{equation}\label{4} 
D^{(T)}= \frac{\sum_{s_i\in S_m}D_i^{(T)}} {m}                                   
\end{equation}

\section{Scheduling Capacity}

To find a robust, efficient scheduling policy, we need to consider scheduling capacity of the FC. Scheduling capacity corresponds to the maximum number of nodes which can be scheduled by the FC in one TS. Since FC has $k$ orthogonal channels, the scheduling capacity of the FC is $k$. If the amount of average harvested energy is so high that the scheduling capacity is exceeded, no 100\% efficient policy exists and energy will keep accumulating (there is an energy surplus). Considering finite batteries, this will cause overflow in the batteries of nodes. Theorem 1
explores the region of energy harvest rates such that a 100\% efficient policy is feasible. 

We shall now make some definitions that will be used in the rest of this section and the paper. We denote by $V_i^{(T)}(t)$ and $V_{av}^{(T)}(t)$, the energy available to sensor $s_i$ in the interval $(T,N]$ and the average of this over all sensors in this time interval, respectively. Also, we denote by $V_{tot}^{(T)}(t)$ the total energy available $(T,N]$: 
\begin{equation}\label{5}
V_i^{(T)}(t)= \left\lfloor B_i(T)+\sum_{t=T+1}^N E_i^h(t)\right\rfloor
\end{equation}   
\begin{equation}\label{6} 
V_{tot}^{(T)}(t)=\sum_{s_i\in S_m}V_i^{(T)}		
\end{equation}
\begin{equation}\label{7} 
V_{av}^{(T)}(t)=\frac{\sum_{s_i\in S_m}V_i^{(T)}}{m}		
\end{equation}	
						
\textbf{Theorem 1 (Scheduling Capacity Theorem):} For $0\leq T<N $, 

\textit{
(i)    If $V_{av}^{(T)}(t)>\frac{k(N-T)}{m}$,  all possible policies will have efficiency below 100\% and battery levels of some sensors grow unboundedly (in practice, considering finite batteries, they will overflow.)}

\textit{
(ii)	If $V_{av}^{(T)}(t)\leq\frac{k(N-T)}{m}$, a 100\% efficient policy that maximizes throughput while keeping battery levels of all sensors finite, exists.}

\textbf{\textit{Proof of Theorem 1:}}

\textit{(i) }In this case,
\begin{equation}\label{10}
 V_{tot}^{(T)}(t)=mV_{av}^{(T)}(t)>{k(N-T)}
\end{equation}

As the total uplink rate available is k data packets per slot, FC can accumulate at most $k(N-T)$ packets from the nodes in the interval $(T,N]$. Suppose that there is a policy $\pi^*$ which can achieve up to scheduling capacity. Then, efficiency of $\pi^*$ equals to the maximum efficiency in the conditions (\ref{10}), and it is represented as below:
\begin{equation}\label{11}
\eta_{max}=\eta_* =\frac{min\left\lbrace k(N-T),V_{tot}^{(T)}(t)\right\rbrace}{V_{tot}^{(T)}(t)}=\frac{k(N-T)}{V_{tot}^{(T)}(t)}  
\end{equation}

If (\ref{10}) is satisfied, the scheduling capacity is exceeded in the interval $(T,N]$.  By (\ref{10}), $\eta_{max}=\eta_*<1$. Hence, there is no 100\% efficient policy which lets FC receive $V_{tot}^{(T)}(t)$ packets from the nodes. By definition of $D^{(T)}$, (\ref{10}) is equivalent to $D^{(T)}>1$. 
Define excess energy as $B_{ex}(t)=\sum_{i=1}^m\left\lfloor B_i(t)\right\rfloor =min\left\lbrace 0,V_{tot}^{(T)}(t)-k(N-T)\right\rbrace$. By definition of $D^{(T)}$, $B_{ex}(t)=k(N-T)(D^{(T)}-1)$.
$D^{(T)}>1$, and $N\rightarrow\infty$, 
battery levels of some sensors grow unboundedly.

\textit{(ii) }In this case, 
\begin{equation}\label{14}
 V_{tot}^{(T)}(t)=mV_{av}^{(T)}(t)\leq{k(N-T)}
\end{equation}
FC can receive maximum $k(N-T)$ data packets from the nodes in the interval $(T,N]$. An omniscient policy could fill up all channels in all time slots as long as there is a sensor with available energy. Trivially, this achieves 100\% efficiency if $D^{(T)}\leq 1$ (equivalent to \ref{14}). It is summarized below:
\begin{equation}\label{15}
\eta_{max}=\eta_* =\frac{min\left\lbrace k(N-T),V_{tot}^{(T)}(t)\right\rbrace}{V_{tot}^{(T)}(t)}=\frac{V_{tot}^{(T)}(t)}{V_{tot}^{(T)}(t)}=1 
\end{equation}
By (\ref{15}), $\eta_{max}=\eta_*=1$. Hence, there is a fully efficient (100\% efficient) policy which makes FC receive $V_{tot}^{(T)}(t)$ packets. Battery levels of all sensors are kept finite.
Hence when $D^{(T)}\leq 1$, there is an optimal policy which is 100\% efficient.

\section{Efficiency of RR-based Policies}

The scheduling problem in this paper are also studied in \cite{18} and \cite{20} for certain specific cases. Both papers propose RR-based policies with quantum=1 which are myopic policies. Then, they prove the optimality of these policies under certain specific cases.  

First, we will investigate the efficiency of RR-based policies by Theorem 2. Then, we will prove by Theorem 3 that there is only a slight difference between the efficiencies of any two RR-based policies in long problem horizon $N(\frac{m}{k}=p<<N)$. Hence, the efficiency of RR-based myopic policies in \cite{18} and \cite{20} are investigated. It is shown that the policies in \cite{18} and \cite{20} are generally suboptimal. 

For the cases that each node has a battery larger than unit size, there is no known myopic policy in the literature. Therefore, we will compare UROP only with the policies in \cite{18} and \cite{20}, and the optimal policy in this paper.

\textbf{Theorem 2:}
\textit{Suppose that $N>>p=\frac{m}{k}\in Z$ and $V_{av}^{(T)}(t)\leq \frac{k(N-T)}{m}$. If there are some sensors $s_i\in S_m$ such that $V_i^{(T)}(t)>\frac{k(N-T)}{m}$, all RR-based policies with quantum=1 will have efficiency below 100\% although a fully efficient policy ($\pi^{fe}$) exists. Moreover, batteries of some sensors will overflow.}

\textbf{\textit{Proof of Theorem 2:}} In this proof, what is implied by RR policy is RR-based policies with quantum=1. We investigate efficiency of RR in the two possible cases:

i. 	If $\sigma=\frac{kN}{m}=\frac{N}{p}\in Z$, RR allocates each node $\sigma$ TSs for transmission.

ii. If $\sigma=\frac{kN}{m}=\frac{N}{p}\notin Z$, RR allocates some nodes $\left\lfloor \sigma \right\rfloor+1$ TSs and other nodes $\left\lfloor \sigma \right\rfloor$ TSs.

Assume that there are some nodes $s_i\in S_m$ such that $V_i^{(T)}(t)>\frac{k(N-T)}{m}$. We denote by $H$ the set of these sensors. By definition $D_i^{(T)}>1$  for nodes $s_i\in H$.

\textbf {Case i: }If the FC schedules $m$ sensors by RR policy in the problem horizon $N$,  RR policy allocates each node $\sigma=\frac{kN}{m}=\frac{N}{p}$ TSs equally. Although $V_{av}^{(T)}(t)\leq \frac{k(N-T)}{m}$, each sensor $s_i\in H$ can transmit maximum $\sigma $ data but cannot transmit  $V_i^{(T)}(t)-\sigma $ data due to RR policy. On the other hand, each of other sensors $s_i\in S_m-H$ can transmit all $V_i^{(T)}(t)$ packets. 
By analogy with scheduling capacity, the efficiency of a RR policy can be represented as below:

\begin{eqnarray} 
\eta_{RR}&=&\frac{\sum_{s_i\in S_m} min\left\lbrace V_i^{(T)}(t),\sigma \right\rbrace}{\sum_{s_i\in S_m}V_i^{(T)}(t)} \label{19} \\
&=&\frac{\sum_{s_i\in H} min\left\lbrace V_i^{(T)}(t),\sigma \right\rbrace + \sum_{s_i \in S_m-H} min\left\lbrace V_i^{(T)}(t),\sigma \right\rbrace}{\sum_{s_i\in S_m}V_i^{(T)}(t)} \label{20}\\
&=&\frac{\sum_{s_i\in H} \sigma + \sum_{s_i \in S_m-H} V_i^{(T)}(t)}{\sum_{s_i\in S_m}V_i^{(T)}(t)} \label{21}\\
&=&1-\frac{\sum_{s_i\in H} (V_i^{(T)}(t)-\sigma)}{\sum_{s_i\in S_m}V_i^{(T)}(t)}  \label{22}
\end{eqnarray}

Considering the definition of $D_i^{(T)}$, $\eta_{RR}$ can also be represented as below:
\begin{equation}\label{23}
\eta_{RR}=1-\frac{\sum_{s_i\in H} (D_i^{(T)}-1)}{\sum_{s_i\in S_m}D_i^{(T)}}
\end{equation}

Since $D_i^{(T)}>1$ for $s_i\in H$, $\eta_{RR}<1$. Hence, suboptimality of RR policy is proven for the first case although there exists an 100\% efficient policy by Theorem 1.

\textbf{Case ii: }If the FC schedules $m$ sensors by RR policy in the problem horizon $N$, RR policy allocates some nodes $\left\lfloor \sigma\right\rfloor+1$ TSs and other nodes $\left\lfloor \sigma \right\rfloor$ TSs for transmission where $\sigma=\frac{kN}{m}=\frac{N}{p}\notin Z$ and $\sigma^+=\sigma-\left\lfloor \sigma \right\rfloor$. To maximize efficiency of RR policy, we assume that each sensor $s_i\in H$ can transmit maximum $\left\lfloor \sigma \right\rfloor+1$ data. However, each of these sensors cannot transmit  $V_i^{(T)}(t)-\left\lfloor \sigma \right\rfloor-1$ data due to RR policy although $V_{av}^{(T)}(t)\leq \frac{k(N-T)}{m}$. On the other hand, each of other sensors $s_i\in S_m-H$ can transmit all $V_i^{(T)}(t)$ data. By the analogy with scheduling capacity, the efficiency of RR policy can be represented as below:

\begin{eqnarray}
\eta_{RR}&=&\frac{\sum_{s_i\in S_m}min\left\lbrace V_i^{(T)}(t),\sigma \right\rbrace}{\sum_{s_i\in S_m}V_i^{(T)}(t)} \label{24} \\
&=&\frac{\sum_{s_i\in H}min\left\lbrace V_i^{(T)}(t),\sigma \right\rbrace + \sum_{s_i\in S_m-H} min\left\lbrace V_i^{(T)}(t),\sigma \right\rbrace}{\sum_{s_i\in S_m}V_i^{(T)}(t)} \label{25}  \\
&=& \frac{\sum_{s_i\in H}(\left\lfloor \sigma \right\rfloor+1) + \sum_{s_i\in S_m-H} V_i^{(T)}(t)}{\sum_{s_i\in S_m}V_i^{(T)}(t)}  \label{26}  \\
&=&1-\frac{\sum_{s_i\in H}(V_i^{(T)}(t)-\left\lfloor \sigma \right\rfloor-1)}{\sum_{s_i\in S_m}V_i^{(T)}(t)} \label{27}
\end{eqnarray}

Considering the definition of $D_i^{(T)}$, $\eta_{RR}$ can also be represented as below:
\begin{eqnarray}
\eta_{RR}&=&1-\frac{\sum_{s_i\in H} (D_i^{(T)}\sigma-\left\lfloor \sigma \right\rfloor-1)}{\sum_{s_i\in S_m}D_i^{(T)}\sigma}  
\label{28} \\
&=&1-\frac{\sum_{s_i\in H} (D_i^{(T)}-1)\sigma-(1-\sigma^+))}{\sum_{s_i\in S_m}D_i^{(T)}\sigma}    \label{29}       
\end{eqnarray}

It is known that $D_i^{(T)}>1$ for $s_i\in H$ and $(1-\sigma^+)<(D_i^{(T)}-1)\sigma$ since $\sigma>>1>1-\sigma^+$. Therefore, $\eta_{RR}<1$. Hence, suboptimality of RR policy is also proven for the second case although there exists an 100\% efficient policy by Theorem 1. From (\ref{23}), (\ref{29}) and Theorem 1, efficiency of RR be as low as $\frac{k}{m}$. This worst case efficiency of $\frac{k}{m}$ occurs when $k$ of the nodes always have sufficient energy to transmit a packet in each TS and the remaining ones have no energy.

For a sufficiently long problem horizons, these results can be extended to RR-based policies with larger quanta. The following remark, used in the rest of the paper, is a consequence of  the assumption there is no battery leakage. 

\textbf{Remark 1 (No battery leakage):} \textit{Let $T_1,T_2\in (0,N]$ and $T_1<T_2$. If $s_i$ is not scheduled (selected) in interval $(T_1,T_2]$, $B_i(T_1)\leq B_i(T_2)$ where $B_i(t)$ is the energy remaining in battery of sensor $s_i$ at the end of TS $t$. That is, $B_i(t)$ does not decrease unless $s_i$ transmits data.}

Theorem 2 states that RR-based policies become suboptimal when $D_i^{(T)}>1$ even for one node. The myopic policies (MP) in \cite{18} and \cite{20} are RR policies with quantum=1. In Theorem 3, it is shown that the MPs have almost same efficiency as any other RR policy with quantum=1.

\textbf{Theorem 3 (Upper and lower bounds on RR throughput ):}
\textit{Assume that $\frac{m}{k}\in Z$. In problem horizon $N$, }
\begin{equation}\label{101}
max\left\{ V_{tot}^{RR}(N)\right\} - min \left\{V_{tot}^{RR}(N)\right\} \leq {m-k}
\end{equation}
\textit{where $ min \left\{V_{tot}^{RR}(N)\right\} $ and $max\left\{ V_{tot}^{RR}(N)\right\}$ are the minimum and maximum throughput which can be achieved under a RR policy with quantum=1, respectively.}

\textbf{\textit{Proof of Theorem 3:}} There are three cases for the problem horizon $N$: 1) $N<\frac{m}{k}$, 2)$N \geq \frac{m}{k}$ and $\frac{kN}{m}\in Z$, and 3) $N \geq\frac{m}{k}$ and $\frac{kN}{m}\notin Z$

\textbf{Case 1:} If  $N<\frac{m}{k}=p$, $N \leq p-1$. Since $min \left\{V_{tot}^{RR}(N)\right\} \geq 0$  and $max \left\{V_{tot}^{RR}(N)\right\} \leq {kN} \leq k(p-1)= m-k$, $max\left\{ V_{tot}^{RR}(N)\right\} - min \left\{V_{tot}^{RR}(N)\right\} \leq {m-k}$. This proves the statement for this case.

\textbf{Case 2:} Denote by $U_i$ the nodes scheduled in TS $i$ where $i\leq p$ and $S_m=\bigcup_{i=1}^p U_i$ . All RR policies have same length period $p$. Denoted by $\tau_l^{RR}$ the $l^{th}$ period of RR, namely, $\tau_l^{RR}=\left[(l-1)p+1,lp \right]$. Assume that $T_1,T_2 \in \tau_j^{RR}$ and $T_1<T_2$ where $T_1$ and $T_2$ are the TSs when a node $s_j$ is scheduled $l^{th}$  time by the FC under two different myopic policies, $\pi^{RR1}$ and $\pi^{RR2}$, respectively. 

By Remark 1, efficiency of $\pi^{RR2}$ in $T_2$ is not lower than that of $\pi^{RR1}$ in $T_1$ for the node $s_j$ since $\pi^{RR2}$ schedules the node later than $\pi^{RR1}$ does. By Remark 1, if a node $s_j$ cannot send data in $T_1$ and can send in $T_2$, then it would certainly have data to send when it is scheduled in $T_1+p$ instead of $T_2$. Therefore, $V_j^{RR1}(T_1)\leq V_j^{RR2}(T_2)\leq V_j^{RR1}(T_1+p)$ for $\forall s_j\in S_m$. 

This means that giving each node one more TS, any RR policy can achieve maximum throughput achieved by most efficient RR. In other words, the least efficient RR can achieve the throughput of the most efficient RR by continuing only one period more. Note that since $U_p$ is the nodes scheduled last under a RR policy, they achieve maximum throughput which can be achieved under RR policy by Remark 1. Therefore, the least efficient RR uses only $m-k$ TSs more than other RR policies to guarantee same throughput. By using the extra $m-k$ TSs which the least efficient RR used, the most efficient RR policy can have throughput $m-k$ more than it has. By giving an example for this situation, Theorem 3 will be proved for this case. Considering the last period $\left[T+1,T+p\right]$, the worst performance of RR occurs when the set of nodes $U_i$ can transmit no data in TS $T+i$; however, they get ready for transmission($B_j(t)\geq 1$) in TS $T+i+1$. Since there is no next TS for $U_p$, nodes of $U_p$ cannot improve their battery states. Therefore, the throughput difference is determined by nodes $s_j\in {S_m-U_p} $. Since $\left|S_m-U_p \right|=m-k$, the difference is $m-k$. This concludes the proof.

\textbf{Case 3:} Assume that $N=sp+c$ where $0<c<p$, $s\in Z$. In case 2, it is shown that the maximum difference is $m-k$ in TS $sp$. In the interval $[sp+1,sp+c]$, $S_c=\bigcup_{i=1}^c U_i$  is scheduled. For the nodes $s_j\in U_i\subset S_c$ , $B_j (t)\geq 1$ in TS $sp$. If  $B_j (sp+i)\geq 2$ $\forall s_j\in U_i\subset S_c$, the throughput difference remains as $m-k$. Unless $B_j (sp+i)\geq 2$ $\forall s_j\in U_i\subset S_c$, the difference remains same or decreases depending on other nodes $s_j\in S_m-S_c$. Hence, it is proved.

\section{Optimal Omniscient Policies}

For the EH scheduling problem, \cite{18},\cite{20} propose RR-based Myopic Policy (MP) and prove that the MP is optimal for certain specific cases. However, Theorem 2 and Theorem 3 state that RR-based policies with quantum=1 (including the MP in \cite{18},\cite{20}) become suboptimal when $D_i^{(T)} > 1$  for some sensor $i$  although an 100\% efficient policy exists ($D^{(T)} \leq 1$). The EH scheduling problem resembles a simplified unicast switch scheduling problem. Like unicast switch scheduling problems, this problem has input queues (energy queues) and feasible activation sets are such that at most $k$ users are scheduled. Different from usual unicast switch scheduling problem setups, buffer (battery) states are not known in this problem; therefore, switch scheduling policies that assume the availability of state information cannot be applied directly. However, these provide intuition for finding an \emph{omniscient scheduling} policy (i.e. one which knows the current battery states) for the EH scheduling problem.

For unicast switch scheduling problems, the following approaches are well-known: Maximum Size Matching (MSM) and Maximum Weight Matching (MWM) \cite{23}, \cite{24}. Maximum Size Matching selects in each TS an activation set with the max number of nonempty queues. On the contrary, Maximum Weight Matching also respects queue size not only whether queues are empty or not. Maximum Size Matching may sometimes cause starvation due to head-of-line (HOL) blocking which limits its throughput to below 100\% in some cases  \cite{23} and \cite{25}. On the other hand, Maximum Weight Matching policies always guarantee 100\% throughput for all admissible traffic (with analogy, $D^{(T)} \leq 1$ in our problem) and two MWM algorithms are offered to achieve 100\% throughput in \cite{23}. However, due to lower computational complexity, MSM policies are sometimes preferred \cite{26}.

Different from unicast switch scheduling problems, there is no preffered output for packets in the EH scheduling problem. All $k$ lines correspond to the same output port  to which any packet can be sent. This implies that HOL blocking does not occur in the EH scheduling problem, so both MSM and MWM will provide 100\% throughput in our problem. Due to lower computational complexity, MSM is preferable. To find an omniscient policy for the EH scheduling problem, we assume that FC knows whether each node can transmit data or not in any TS. With this knowledge, there is no unique optimal omniscient policy for this problem. We shall concentrate on one optimal policy which provides intuition to find a near optimal, nonomniscient, online policy later. 

To find such a policy, we map the problem onto a variation of block-packing game Tetris. A different Tetris model which we are inspired by was previously used in multicast switch scheduling problems \cite{27} and \cite{28}. In this model, packets from same input are sent to different output ports. In our case, different from the Tetris model of \cite{27} and \cite{28}, the packets from same input are sent to same output port if the input is scheduled to transmit data to that output port in our model (The model is shown in Figure 2.). That is the critical point which provide us intuition to find a near optimum, nonomniscient, online policy in the next section.

\begin{figure}
	\centering	\includegraphics[width=0.80\textwidth]{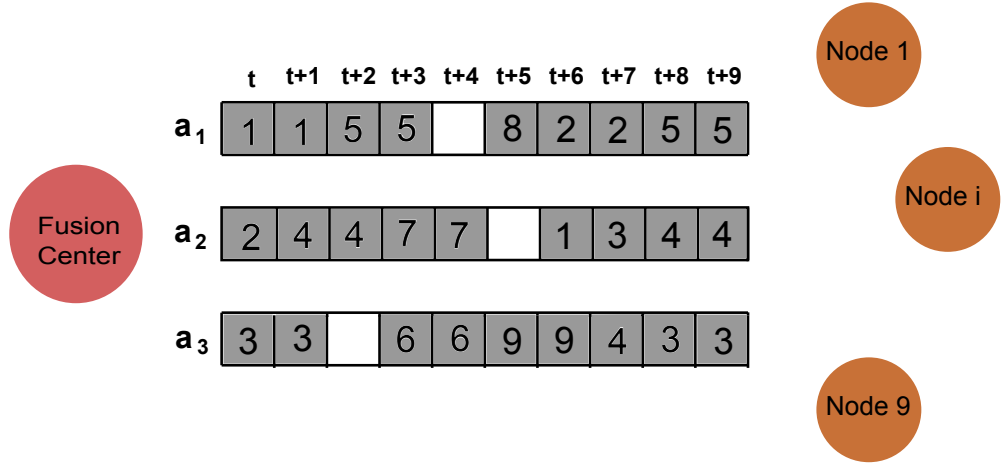}
	\caption{An example scheduling table kept by the fusion center (FC) for $m=9$, $k=3$ during the interval $\left[t, t+9\right]$. Dark colored TSs represent busy slots labeled by node ID using the slot, and the white ones represent idle ones. 3 of 30 slots are idle even under an optimum omniscient policy (UP) ($D^{(T)}=0.9$). UP allocates the slots in order to leave the least number of slots idle: resembling a Tetris game. Note that each node can use only one channel at a time.}
	\label{Figure 2}
\end{figure} 

Based on our tetris model, we propose an omniscient, optimum scheduling policy, \emph{Uniformizing Policy (UP)} for all admissible EH process. Considering nonuniform EH processes at all $m$ nodes, UP uses the empty output ports to schedule the nodes in each TS. If there are some nodes which are scheduled in previous TS but does not have enough energy to transmit data in current TS, UP schedules new nodes. By scheduling new nodes, UP prevents output ports to remain idle and balances the load in each of $k$ output ports. Hence, UP \emph{uniformizes} the nonuniform EH processes of $m$ nodes such that all packets are scheduled in each of $k$ output port almost equally. By this almost equal partition of the packets sent by nodes, UP makes \emph{uniformization} and provides 100\% throughput under all admissible uniform and non-uniform EH processes. 
\vspace{0.2 in}

The operation of UP is summarized below:

1.	Order the nodes arbitrarily and use this order throughout problem horizon.

2.	Schedule the first $k$ nodes in the ordering that have enough energy to transmit a packet.

3.	At the beginning of the next TS, check the $k$ nodes that were just scheduled. Replace those without energy to transmit a packet with new ones, respecting the initial order. If less than $k$ nodes with enough energy can be found, schedules those nodes only.

4.	Continue in a cyclic way.


\section{A Near-Optimal Online Policy}

\subsection{Uniformizing Random Ordering Policy (UROP)}

Assuming that all EH process is known in previous section, an optimal omniscient solution is proposed for the EH scheduling problem. However, the battery states of the nodes are not known in the exact EH scheduling problem. Therefore, we propose a near-optimal online scheduling policy by using Lemma 1 (stated below) for all admissible EH processes $D^{(T)}<1$. $D^{(T)}<1$ means that there exists always idle TSs over a problem horizon even if an optimal policy is applied. Lemma 1 states that if a scheduled node cannot transmit data in TS $t$, an 100\% efficient policy is applied to that node until TS $t$. Therefore, we propose UROP which uses the idle TSs to determine battery state of the scheduled nodes (whether a node has enough energy to transmit data or not). 

Since EH processes are completely unpredictable for some EH sources \cite{1,3}, UROP orders the nodes randomly before starting to schedule them. 
\vspace{0.2 in}

UROP operates as below:

1. Schedule the first $k$ nodes according to initially determined random order. 

2. If a scheduled node transmits data to FC in that TS, then it continues to be scheduled. 

3. Otherwise, FC starts to schedule the nodes which have highest priority in the cyclic random order instead of the leaving ones. 
\vspace{0.2 in}

To schedule all nodes once, FC uses $m$ nodes to complete a period (all nodes are scheduled once). As $D\rightarrow1$, the ratio of idle TSs over whole problem horizon decreases. The algorithm, UROP, whose operation is described above is hence an adaptive and near optimal policy. In this section, the efficiency of UROP is investigated by assuming that no node behaves as \emph{an elephant node} (defined below). In section VII, it is shown that UROP is asymptotically optimal over infinite horizon for all admissible EH processes.

\textit{Definition 1 (Elephant node):} If the node who is next in line for selection by the FC happens to be already transmitting continuously since its last selection, the node is said to behave as an \emph{elephant node} between the previous selection (scheduling) time and the current selection time. In this case, FC selects the next node to schedule for one of the empty channels and the elephant node continues to transmit on its assigned channel as before. Figure 3 represents an elephant node.
\begin{figure}
	\centering	\includegraphics[width=0.70\textwidth]{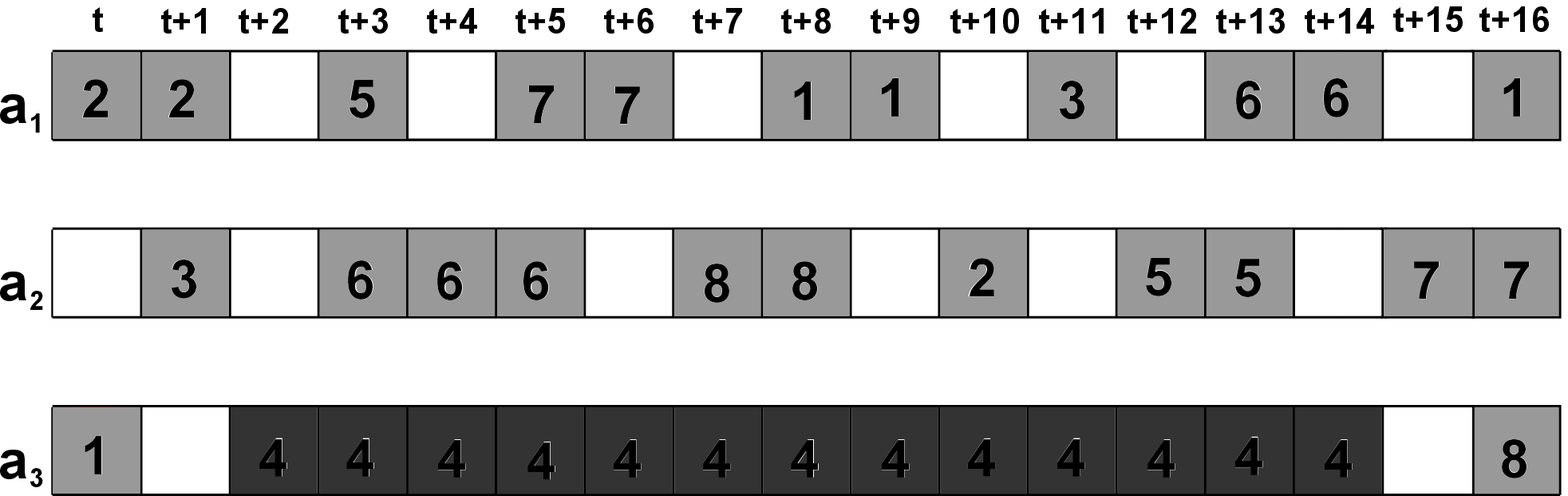}
	\caption{An example scheduling table kept by the fusion center (FC) for $m=8$,$k=3$ during the interval $\left[t, t+16\right]$. Dark colored TSs represent busy slots labeled by node ID, and the white ones represent idle ones. Node 4 behaves as an elephant node since it does not idle within a round continues transmission from $t+2$, until $t+14$. Note that it has already transmitted data in $t+12$ when it is next supposed to be scheduled.}
	\label{Figure 3}
\end{figure}

\subsection{Efficiency of UROP in Finite Horizon Case}

In this part, the efficiency of UROP is investigated in general case of EH process. First, several lemmas are stated and proved. Then, they are used to prove Theorem 4 and Theorem 5.

\textbf{Lemma 1 (Partial Optimality): }\textit{If $B_i(t)<1$ for a sensor $s_i$ at the end of TS $t$, an optimal policy has been applied for sensor $s_i$ and efficiency is 100\% for sensor $s_i$ up to $t$.}

\textbf{\textit{Proof of Lemma 1:}} The number of data packets which could be sent by sensor $s_i$ with the remaining energy in TS $t$ is $\left\lfloor B_i(t) \right\rfloor $. Since $B_i (t)<1$, $\left\lfloor B_i(t) \right\rfloor=0 $ . By TS $t$, $s_i$ has transmit all data which could be sent with $E_i^{tot}(t)$, and efficiency is 100\% for node $s_i$ until TS $t$.

\textbf{Remark 2:}\textit{ If $E_i^{tot}(t)$  is the total amount of harvested energy in sensor $s_i$ until TS $t$ and  $V_i^{opt}(t)$ is the number of packets (throughput) which could be sent by sensor $s_i$ until TS $t$ under $\pi^{opt}$, $V_i^{opt}(t)=\left\lfloor E_i^{tot}(t)\right\rfloor$. Recall that $\pi^{opt}=\pi^{fe}$ for $0\leq D \leq 1$}.


 

Now, we will define some new parameters which will be used in Lemma 2, Lemma 3, Lemma 4 and Theorem 4. Remember that $A_k=\left\{a_1,a_2,..,a_k\right\}$ is the set of mutually orthogonal channels of FC. $\gamma_l^j$ is the $l^{th}$ idle TS for $a_j$, the $j^{th}$ channel of the FC. In this TS $\gamma_l^j$, FC drops a node using  $j^{th}$ channel and start to schedule another node in same channel. $\forall \gamma_l^j \in T_I^{(j)}$ Let's denote by $T_I^{(j)}$ the idle TSs for $a_j$, the $j^{th}$ channel of the FC. In these TSs, FC drops some of the $k$ nodes and starts to schedule other nodes in their place. $T_I$ is the set which consists of all pairs $(a_j,\gamma_l^j)$ where $a_j\in A_k$ ,$\gamma_l^j \in T_I^{(j)}$. Figure 4 represents the pairs (idle TSs in 2-dimension).

Let's denote by $\xi_i^{(f)}$ and $\xi_i^{(f-1)}$ the idle TSs when FC starts to schedule node $s_i$ for the last time and for the second last time, respectively. If $a_u\in A_k$ and $\gamma_v^u\in T_I^{(u)}$, $F_1$ and $F_2$ are the set of all pairs $\left(a_u,\gamma_v^u\right)$ such that $\gamma_v^u=\xi_i^{(f)}$ for a $s_i\in S_m$ and the set of all pairs $\left(a_u,\gamma_v^u\right)$ such that $\gamma_v^u=\xi_i^{(f-1)}$ for a $s_i\in S_m$. There are $m$ nodes so $\left| F_1\right|=\left|F_2\right|=m$. If $a_p\in A_k$ and $\gamma_q^p\in T_I^{(p)}$, $G_1$ is the set of all pairs $(a_p,\gamma_q^p)$  such that $\gamma_q^p\neq \xi_i^{(f)}$ for $s_i\in S_m$. Moreover, $G_2$ is the set of all pairs $(a_p,\gamma_q^p)$  such that $\gamma_q^p\neq \xi_i^{(f)}$ and $\gamma_q^p\neq \xi_i^{(f-1)}$ for $s_i\in S_m$. In other words, $G_1=T_I-F_1$ and $G_2=T_I-(F_1 \cup F_2)$. 

\begin{figure}
	\centering	\includegraphics[width=0.60\textwidth]{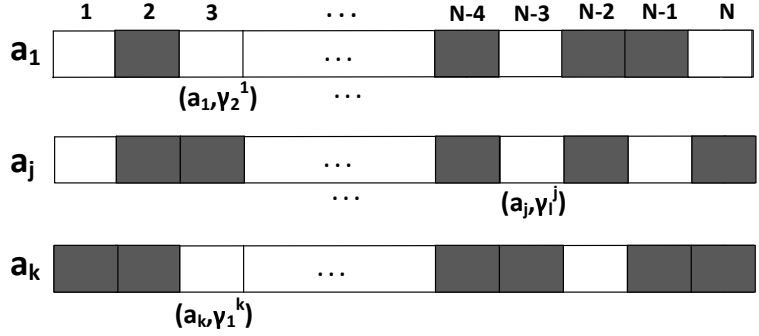}
	\caption{An example scheduling table kept by the fusion center (FC) for all $k$ channels over problem horizon $N$. Dark colored TSs represent busy slots, and the white ones represent idle ones.}
	\label{Figure 4}
\end{figure}

\textbf{Lemma 2:} \textit{If $\left(a_u,\gamma_v^u\right) \in (F_1 \cup F_2)$},

\textit{i. There exists no $(a_p,\gamma_q^p)\in G_1$  such that $\gamma_q^p\neq \gamma_v^u$ for $\exists \left(a_u,\gamma_v^u\right) \in F_1$}

\textit{ii. There exists no $(a_p,\gamma_q^p)\in G_2$  such that $\gamma_q^p\neq \gamma_v^u$ for $\exists \left(a_u,\gamma_v^u\right) \in (F_1 \cup F_2)$}

\textbf{\textit{Proof of Lemma 2:}}

\textbf{\textit{Part i:}}
Assume that there is a pair $(a_p,\gamma_q^p)\in G_1$ such that $\gamma_q^p\neq \gamma_v^u$ for $\exists \left(a_u,\gamma_v^u\right)\in F_1$. Since $\gamma_q^p\neq \xi_i^{(f)}$ for $\forall s_i\in S_m$, the node $s_r$ which is selected by the FC in TS $\xi_q^p$ will be selected by the FC at least once more ($\gamma_q^p<\xi_r^{(f)}$). According to UROP, a node $s_r$ which is selected in TS $T_1$ cannot be selected by the FC in TS $T_2$ unless $\forall s_i\in S_m-{s_r}$ are selected in the interval$[T_1,T_2]$.  Since $\gamma_q^p>\gamma_v^u=\xi_i^{(f)}$ for $\exists s_i$ , these nodes cannot be selected by the FC in $[\gamma_q^p,\xi_j^{(f)}]$. Therefore, there exists no $(a_p,\gamma_q^p)\in G_1$  such that $\gamma_q^p\neq \gamma_v^u$ for $\exists \left(a_u,\gamma_v^u\right) \in F_1$

\textbf{\textit{Part ii:}}
 $G_1=T_I-F_1$ and $G_2=T_I-(F_1\cup F_2)=(T_I-F_1)-F_2$. Replacing $T_I-F_1$ and $F_2$ with $T_I$ and $F_1$ , respectively, in case 1, we can said that there exists no $(a_p,\gamma_q^p)\in G_2$ such that $\gamma_q^p>\gamma_v^u$ for $\exists (a_u,\gamma_v^u)\in F_2$. By case 1 of lemma 2, there exists no $(a_p,\gamma_q^p)\in F_2$ such that $\gamma_q^p>\gamma_v^u$ for $\exists (a_u,\gamma_v^u)\in F_1$. Therefore, there exists no $(a_p,\gamma_q^p)\in G_2$  such that $\gamma_q^p\neq \gamma_v^u$ for $\exists \left(a_u,\gamma_v^u\right) \in (F_1 \cup F_2)$.

\textbf{Lemma 3:}\textit{ If $\zeta_j^{(f)}$ is the idle TS when FC stops to schedule node $s_j$ for the last time and $L$ is the set of the idle TSs $\zeta_j^{(f)}$, $L\subset(F_1\cup F_2)$.}

\textbf{\textit{Proof of Lemma 3:}}
Recall that $(F_1\cup F_2)\subseteq T_I$. It can be said that FC starts to schedule a node iff it leaves (stops to schedule) another node. $(F_1\cup F_2)$ includes two consecutive time (the last and second last time) when FC starts to schedule a node for all nodes.  Assume that FC schedules a node $s_i$. Unless FC stops to schedule the node $s_i$, it cannot start to schedule the node $s_i$ again. Therefore, $(F_1\cup F_2)$ includes at least one departure time for each node.  Since $(F_1\cup F_2)$ includes the latest $2m$ idle TSs and at least one departure time for each node, $\zeta_j^{(f)}\in (F_1\cup F_2)$ for $\forall s_j$. Hence, $L\subset(F_1\cup F_2)$.

Now, we write Lemma 4 and Lemma 5 which will help us find the lower bound of efficiency.

\textbf{Lemma 4: }\textit{Assume that $S_k\subset S_m$  is the set of $k$ nodes which are scheduled last by the FC in problem horizon $N$, each node $s_j\in S_k$  transmits $\left(N-\xi_j^{(f)}\right)$ data in the interval $\left[\xi_j^{(f)}+1,N\right]$.}

\textbf{\textit{Proof of lemma 4:}} If $S_k\subset S_m$ is the set of last $k$ nodes which are scheduled by the FC in problem horizon $N$, there will be no other selection so no idle TS until deadline of problem horizon $N$. Since each node can transmit at most one packet in each TS, each node $s_j\in S_k$ transmits $\left(N-\xi_j^{(f)}\right)\times 1= \left(N-\xi_j^{(f)}\right)$ data in the interval $\left[\xi_j^{(f)}+1,N\right]$.

\textbf{Lemma 5:} \textit{Assume that $T_1,T_2\in \left(0,N\right]$ and $T_1<T_2$. If $s_i$ is not scheduled in interval $[T_1,T_2]$, $V_i^{opt}(T_1)\leq V_i^{opt}(T_2)$ where $V_i^{opt}(t)$ is the number of packets (throughput) which could be sent by sensor $s_i$ until TS $t$ under optimal policy $\pi^{opt}$.}

\textbf{\textit{Proof of lemma 5:}} By Remark 2, $V_i^{opt}(T_1)$ and $V_i^{opt}(T_2)$ can be written as below:
\begin{equation}\label{30}
V_i^{opt}(T_1)=\left\lfloor E_i^{tot}(T_1)\right\rfloor and V_i^{opt}(T_2)=\left\lfloor E_i^{tot}(T_2)\right\rfloor
\end{equation}

From Remark 1 (No battery leakage), the inequality below is written for $\forall s_i\in S_m$
\begin{equation}\label{31}
	E_i^{tot}(T_1)\leq E_i^{tot}(T_2) 
\end{equation}
\begin{equation}\label{32}
\left\lfloor E_i^{tot}(T_1)\right\rfloor+\left(E_i^{tot}(T_1)\right)^+\leq \left\lfloor E_i^{tot}(T_2)\right\rfloor+\left(E_i^{tot}(T_2)\right)^+
\end{equation}

By putting (\ref{30}) into (\ref{32}),
\begin{equation}\label{33}
V_i^{opt}(T_1)+\left(E_i^{tot}(T_1)\right)^+\leq V_i^{opt}(T_2)+\left(E_i^{tot}(T_2)\right)^+
\end{equation}

By using (\ref{33}), it is shown that $V_i^{opt}(T_1)\leq V_i^{opt}(T_2)$ is possible; however, $V_i^{opt}(T_1)> V_i^{opt}(T_2)$ is not possible. There are three cases as below:

\textit{i)}$V_i^{opt}(T_1)= V_i^{opt}(T_2)\Rightarrow \left(E_i^{tot}(T_1)\right)^+\leq \left(E_i^{tot}(T_2)\right)^+$ since $E_i^{tot}(T_1)\leq E_i^{tot}(T_2)$ 

\textit{ii)}$V_i^{opt}(T_1)< V_i^{opt}(T_2)\Rightarrow E_i^{tot}(T_1)< E_i^{tot}(T_2)$ since $0\leq \left(E_i^{tot}(T_1)\right)^+$, $\left(E_i^{tot}(T_2)\right)^+<1$ and $V_i^{opt}(T_1), V_i^{opt}(T_2)\in Z$ 

\textit{iii)}$V_i^{opt}(T_1)> V_i^{opt}(T_2)\Rightarrow E_i^{tot}(T_1)> E_i^{tot}(T_2)$ since $0\leq \left(E_i^{tot}(T_1)\right)^+$, $\left(E_i^{tot}(T_2)\right)^+<1$ and $V_i^{opt}(T_1), V_i^{opt}(T_2)\in Z$ This situation contradicts with (\ref{32}). 

Hence,$E_i^{tot}(T_1)\leq E_i^{tot}(T_2)$ for $\forall s_i \in S_m$ and $V_i^{opt}(T_1)\leq V_i^{opt}(T_2)$ for $\forall s_i \in S_m$.

\textbf{Theorem 4 (Efficiency Bounds of UROP): }\textit{Last departure time of the node $s_j=s_0$ which satisfies $\zeta_j^{(f)}\leq \zeta_i^{(f)}$ for $\forall s_i\in S_m-\left\{s_j\right\}$  is denoted by $\zeta_j^{(f)}=T_0$. In problem horizon $N$, the efficiency of UROP is bounded as below:} 
\begin{equation}\label{102}
1-\frac{k(N-T_0)}{\sum_{i=1}^m V_i^{opt}(N)}\leq \eta_{UROP}\leq 1
\end{equation}
\textit{where $V_i^{opt}(N)$ is the number of packets (throughput) which could be transmitted by sensor $s_i$ until TS $t$ (included) under optimal policy $\pi^{opt}$.}

\textbf{\textit{Proof of Theorem 4: }}$V_i(t)$ is the number of packets (throughput) which have been sent by sensor $s_i$ until TS $t$. $V_i^{(f)}\left(\zeta_i^{(f)}\right)$ is the number of packets (throughput) which could be transmitted by sensor $s_i$ in the interval $\left[ \zeta_i^{(f)}, N\right]$.

$V_i^{opt}=V_i\left(\zeta_i^{(f)}\right)+V_i^{(f)}\left(\zeta_i^{(f)}\right)$ can be written for $\forall s_i\in S_m$. By Lemma 1, $V_i\left(\zeta_i^{(f)}\right)$ is the throughput in TS $\zeta_i^{(f)}$ until when an optimum policy $\pi^{opt}$ is applied to node $s_i$. Therefore, $V_i^{(f)}\left(\zeta_i^{(f)}\right)$ is the only factor for throughput loss of node $s_i\in S_m-S_k$. For $s_i\in S_k$, the throughput loss by $V_i^{(f)}\left(\zeta_i^{(f)}\right)$ decreases by $\left(N-\xi_i^{(f)}\right)$ by Lemma 4.

Hence, the efficiency of UROP in problem horizon $N$ can be written as below:  
\begin{eqnarray}
\eta_{UROP}&=&\frac{\sum_{i=1}^m V_i\left(\zeta_i^{(f)}\right)+\sum_{s_i\in S_k}\left(N-\xi_i^{(f)}\right)}{\sum_{i=1}^m V_i^{opt}(N)} \nonumber \\
&=& \frac{\sum_{i=1}^m V_i\left(\zeta_i^{(f)}\right)+ \sum_{s_i\in S_k}\left(N-\xi_i^{(f)}\right)} {\sum_{i=1}^m V_i\left(\zeta_i^{(f)}\right)+\sum_{i=1}^m V_i^{(f)}\left(\zeta_i^{(f)}\right)} \label{34}
\end{eqnarray}

By using Lemma 4, the term $\sum_{s_i\in S_k}\left(N-\xi_i^{(f)}\right)$ is added to numerator in (\ref{34}) since $s_i\in S_k$ are not considered to be left by the FC in TS $N$. It is assumed that $\zeta_i^{(f)}<N$ for $\forall s_i\in S_k$.

By (\ref{34}), we can upper bound $\eta_{UROP}$.

\textbf{\textit{i. Upper bound for efficiency of UROP}}

Efficiency of a policy cannot be more than 100\% ($\eta\leq 1$). From (\ref{34}), $\eta_{UROP}=1$ only if the equality (\ref{35}) is satisfied:
\begin{equation}\label{35}
\sum_{s_i\in S_k}\left(N-\xi_i^{(f)}\right)=\sum_{i=1}^m V_i^{(f)}\left(\zeta_i^{(f)}\right)
\end{equation}

(\ref{35}) comes true only if (\ref{36}) is satisfied:
\begin{equation}\label{36}
V_i^{(f)}\left(\zeta_i^{(f)}\right) = \left\{ \begin{array}{ll}
0 & \textrm{if $s_i\in S_m-S_k$}\\
\left(N-\xi_i^{(f)}\right) & \textrm{if $s_i\in S_k$}
\end{array} \right.
\end{equation}

If sensors harvest energy such that (\ref{36}) is satisfied, $\eta_{UROP}=1$. Therefore, upper bound of $\eta_{UROP}$ is 100\%, namely, $\eta_{UROP}\leq 1$.

By (\ref{34}), let's find the lower bound of $\eta_{UROP}$.

\textbf{\textit{ii. Lower bound for efficiency of UROP }}

The inequalities below can be written for a long problem horizon $N$. 
\begin{equation}\label{37}
\sum_{s_i\in S_k}\left(N-\xi_j^{(f)}\right)\leq \sum_{i=1}^m V_i^{(f)}\left(\zeta_i^{(f)}\right)<<\sum_{i=1}^m V_i\left(\zeta_j^{(f)}\right)<\sum_{i=1}^m V_i^{opt}(N)
\end{equation}

To find the lower bound of $\eta_{UROP}$, we will define a loss function $V_{loss}$ in (\ref{38}) according to (\ref{34}) and maximize $V_{loss}$ by considering the worst case.
\begin{equation}\label{38}
V_{loss}=\sum_{i=1}^m V_i^{(f)}\left(\zeta_i^{(f)}\right)-\sum_{s_i\in S_k}\left(N-\xi_i^{(f)}\right)
\end{equation}

In (\ref{38}), $V_{loss}$ can be maximized by minimizing $\sum_{s_i\in S_k}\left(N-\xi_i^{(f)}\right)$. 
Since $\xi_i^{(f)}\leq N$ for $\forall s_i\in S_k$,$\sum_{s_i\in S_k}\left(N-\xi_i^{(f)}\right)\geq 0$. This occurs only if $\xi_i^{(f)}= N$ for $\forall s_i\in S_k$.

By Equation (\ref{5}),
$V_i^{(f)}\left(\zeta_i^{(f)}\right)=\left\lfloor B_i\left(\zeta_i^{(f)}\right)+\sum_{t=\zeta_i^{(f)}+1}^N E_i^h(t)\right\rfloor$.

Since $\sum_{s_i\in S_k}\left(N-\xi_i^{(f)}\right)\geq 0$, $V_{loss}$ is maximized if $\sum_{s_i\in S_k}\left(N-\xi_i^{(f)}\right)= 0$. Then (\ref{38}) converts into (\ref{41}).
\begin{equation}\label{41}
V_{loss}=\sum_{s_i\in S_m} V_i^{(f)}\left(\zeta_i^{(f)}\right)
\end{equation}

We denote by $S_k^{(lf)}$ the set of $k$ nodes which satisfy $\zeta_i^{(f)}\leq \zeta_j^{(f)}$ for $\forall s_i\in S_k^{(lf)}$ and $s_i\in S_m-S_k^{(lf)}$. $V_{loss}$ can be written as follows:
\begin{equation}\label{42}
V_{loss}=\sum_{s_i\in S_m} V_i^{(f)}\left(\zeta_i^{(f)}\right)=\sum_{s_i\in S_k^{(lf)}} V_i^{(f)}\left(\zeta_i^{(f)}\right)+\sum_{s_i\in S_m-S_k^{(lf)}} V_i^{(f)}\left(\zeta_i^{(f)}\right) 
\end{equation}

While maximizing $V_{loss}$ , (\ref{43}) must be considered for $\forall s_i\in S_m$ from Theorem 1.
\begin{equation}\label{43}
V_{av}^{(\zeta_i^{(f)})}(t) \leq \frac{k(N-T)}{m}, \forall s_i\in S_m
\end{equation}

From (\ref{43}), FC can accumulate maximum $k$ data. This scheduling capacity can be achieved if there is an energy harvesting process such that $k$ of $m$ nodes can transmit 1 data in each TS and the remaining nodes can transmit no data.

In this case, $V_{loss}$ becomes maximum when each sensor $s_i\in S_k^{(lf)}$ harvests 1 unit energy and the other sensors $s_i\in S_m-S_k^{(lf)}$ harvest almost no energy in each TS.It can be shown as below in (\ref{44}). By putting (\ref{41}) in (\ref{42}), (\ref{44}) can be written as below:
\begin{equation}\label{44}
V_{loss}=\sum_{s_i\in S_k^{(lf)}}\left\lfloor B_i\left(\zeta_i^{(f)}\right)+\sum_{t=\zeta_i^{(f)}+1}^N E_i^h(t)\right\rfloor+\sum_{s_i\in S_m-S_k^{(lf)}}\left\lfloor B_i\left(\zeta_i^{(f)}\right)+\sum_{t=\zeta_i^{(f)}+1}^N E_i^h(t)\right\rfloor
\end{equation}

Last departure time of the node $s_j=s_0$ which satisfies $\zeta_i^{(f)}\leq \zeta_j^{(f)}$ for $\forall s_i\in S_m-\left\{s_j\right\}$ is denoted by $\zeta_j^{(f)}=T_0$. By using Lemma 5, we write an upper bound for $V_{loss}$ as below in (\ref{45}):
\begin{equation}\label{45}
V_{loss}^{'}=\sum_{s_i\in S_k^{(lf)}}\left\lfloor B_i\left(T_0\right)+\sum_{t=T_0+1}^N E_i^h(t)\right\rfloor+\sum_{s_i\in S_m-S_k^{(lf)}}\left\lfloor B_i\left(T_0\right)+\sum_{t=T_0+1}^N E_i^h(t)\right\rfloor \geq V_{loss}
\end{equation}

To maximize $V_{loss}$, maximizing $V_{loss}^{'}$ will be enough so take $V_{loss}=V_{loss}^{'}$. To satisfy this equality, we assume that $T_i=T_0$ for $s_i\in S_k^{(lf)}$. By using (\ref{43}), the inequality (\ref{46}) can be written
\begin{equation}\label{46}
V_{av}^{(T_0)}(t)=\frac{1}{m}\sum_{s_i\in S_m} V_i^{(T_0)}(t) \leq \frac{k(N-T_0)}{m}
\end{equation}
\begin{equation}\label{47}
V_{tot}^{(T_0)}(t)=\sum_{s_i\in S_m} V_i^{(T_0)}(t) \leq k(N-T_0)  
\end{equation}
\begin{equation}\label{48}
V_{tot}^{(T_0)}(t)=\sum_{s_i\in S_k^{(lf)}} V_i^{(T_0)}(t)+\sum_{s_i\in S_m-S_k^{(lf)}} V_i^{(T_0)}(t) \leq k(N-T_0) 
\end{equation}
\begin{eqnarray}\label{49}
V_{tot}^{(T_0)}(t)= \sum_{s_i\in S_k^{(lf)}} \left[V_i^{(T_0)}(t)-V_i^{(f)}\left(\zeta_i^{(f)}\right)+ V_i^{(f)}\left( \zeta_i^{(f)} \right)\right]\nonumber \\ + \sum_{s_i\in S_m-S_k^{(lf)}} \left[V_i^{(T_0)}(t)-V_i^{(f)}\left(\zeta_i^{(f)}\right)+ V_i^{(f)}\left( \zeta_i^{(f)} \right)\right] \leq k(N-T_0) 
\end{eqnarray}

Since $T_0=\zeta_i^{(f)}$ for all $s_i\in S_k^{(lf)}$, $V_i^{(T_0)}(t)-V_i^{(f)}\left(\zeta_i^{(f)}\right)=0$ for all $s_i\in S_k^{(lf)}$. Hence, the inequality converts into (\ref{50}),
\begin{eqnarray}\label{50}
V_{tot}^{(T_0)}(t)= \sum_{s_i\in S_k^{(lf)}} V_i^{(f)}\left( \zeta_i^{(f)} \right)+ \sum_{s_i\in S_m-S_k^{(lf)}} \left[V_i^{(T_0)}(t)-V_i^{(f)}\left(\zeta_i^{(f)}\right)\right] \nonumber \\ + \sum_{s_i\in S_m-S_k^{(lf)}}\left[V_i^{(f)}\left( \zeta_i^{(f)} \right)\right] \leq k(N-T_0)
\end{eqnarray}

By using (\ref{42}) and (\ref{45}) for $V_{loss}^{'}$  in (\ref{50}), (\ref{51}) can be written as below:
\begin{equation}\label{51}
V_{tot}^{(T_0)}(t)= V_{loss}^{'}+ \sum_{s_i\in S_m-S_k^{(lf)}} \left[V_i^{(T_0)}(t)-V_i^{(f)}\left(\zeta_i^{(f)}\right)\right] \leq k(N-T_0)
\end{equation}

Since $T_0\leq \zeta_i^{(f)}$ for all $s_i\in S_m-S_k^{(lf)}$, by Lemma 5, (\ref{52}) can be written
\begin{equation}\label{52}
V_i^{opt}(T_0) \leq V_i^{opt}\left( \zeta_i^{(f)} \right), \forall s_i\in S_m-S_k^{(lf)}
\end{equation}
\begin{equation}\label{53}
V_i^{tot}(N)-V_i^{opt}(T_0) \geq V_i^{opt}(N)-V_i^{opt}\left( \zeta_i^{(f)} \right), \forall s_i\in S_m-S_k^{(lf)}
\end{equation}
\begin{equation}\label{54}
V_i^{(T_0)}(T_0) \geq V_i^{(f)}\left( \zeta_i^{(f)} \right), \forall s_i\in S_m-S_k^{(lf)}
\end{equation}

By (\ref{51}), to maximize $V_{loss}^{'}$, $\sum_{s_i\in S_m-S_k^{(lf)}} \left[V_i^{(T_0)}(t)-V_i^{(f)}\left(\zeta_i^{(f)}\right)\right]$ should be minimized. By (\ref{54}), 
$ V_i^{(T_0)}(t)-V_i^{(f)}\left(\zeta_i^{(f)}\right)\geq 0 $.
If $ V_i^{(T_0)}(t)-V_i^{(f)}\left(\zeta_i^{(f)}\right)= 0 $, (\ref{51}) converts into (\ref{56}),
\begin{equation}\label{56}
V_{tot}^{(T_0)}(t)=V_{loss}^{'}\leq k(N-T_0)
\end{equation}
By using (\ref{45}) and (\ref{56}),
\begin{equation}\label{58}
V_{loss}\leq k(N-T_0)
\end{equation}

By (\ref{34}), efficiency of UROP can be written as below: 
\begin{eqnarray}
\eta_{UROP}=
1-\frac{\sum_{i=1}^m V_i^{(f)}\left(\zeta_i^{(f)}\right)- \sum_{s_i\in S_k}\left(N-\xi_i^{(f)}\right)} {\sum_{i=1}^m V_i\left(\zeta_i^{(f)}\right)+\sum_{i=1}^m V_i^{(f)}\left(\zeta_i^{(f)}\right)}  \label{61}
\end{eqnarray}

Recall that $\sum_{s_i\in S_k}\left(N-\xi_i^{(f)}\right)\geq 0$. From (\ref{37}), $\sum_{i=1}^m V_i^{(f)}\left(\zeta_i^{(f)}\right)<<\sum_{i=1}^m V_i\left(\zeta_i^{(f)}\right)$,
\begin{eqnarray}
\eta_{UROP} &\geq&1-\frac{V_{loss}}{\sum_{i=1}^m V_i^{opt}(N)}  \label{105} \\
&\geq&1-\frac{\sum_{i=1}^m V_i^{(f)}\left(\zeta_i^{(f)}\right)}{\sum_{i=1}^m V_i^{opt}(N)} \label{62}  \\
&\geq&1-\frac{k(N-T_0)}{\sum_{i=1}^m V_i^{opt}(N)} \label{63}
\end{eqnarray}


Hence, Theorem 4 is proven and the efficiency of UROP is bounded as below:
\begin{equation}\label{65}
1-\frac{k(N-T_0)}{\sum_{i=1}^m V_i^{opt}(N)}\leq \eta_{UROP}\leq 1
\end{equation}

\emph{When elephant nodes are present:} Regular nodes scheduled by UROP, give rise to at least one idle TS in a period (frame). However, this does not hold for elephant nodes. If there are nodes that behave as elephant nodes in a period, these do leave any TS empty in that period. Consequently, for these nodes UROP behaves as UP, which does not give up TS to determine the battery states of nodes. Hence, efficiency bounds in Theorem 4 are also valid in case of elephant nodes.

Considering \emph{the worst case} in Theorem 4, we found lower and upper bounds for the efficiency of UROP in terms of parameters. $k$ is known and $V_i^{opt}(N)$ can be found for each node $s_i$ by Remark 2. However, the parameter $T_0$ cannot be determined unless all details of scheduling in problem horizon is known. Due to the incertainty of $T_0$, Theorem 4 does not give sufficient information about efficiency of UROP. As we mentioned in Section II, expected average reward is a suitable performance measure for the EH scheduling policy over finite or infinite horizon \cite{22}. Considering $T_0$ (and the
other departure times of nodes) as ergodic processes depending on EH processes, we take expectation of the bounds of UROP in Theorem 5. Hence, the bounds of UROP can be determined in expected manner.

\textbf{Theorem 5: }\textit{For $0<D<1$, expected efficiency of UROP is bounded as below:}
\begin{equation}\label{103}
1-\frac{2m}{(1-D)DNk}\leq E\left\{\eta_{UROP}\right\}\leq 1
\end{equation}
\textit{where $D,N,m,$ and $k$ are density, problem horizon length, number of the sensors, number of the orthogonal channels of the FC, respectively.}

\textbf{\textit{Proof of Theorem 5: }}By Theorem 4, the efficiency of UROP can be written as below:
\begin{equation}\label{66}
1- \frac{k(N-T_0)}{\sum_{i=1}^m V_i^{opt}(N)}\leq \eta_{UROP}\leq 1
\end{equation}
\begin{equation}\label{67}
1- \frac{k(N-T_0)}{V_{tot}(N)}\leq \eta_{UROP}\leq 1
\end{equation}
\begin{equation}\label{68}
1- E\left\{\frac{k(N-T_0)}{V_{tot}(N)}\right\} \leq E\left\{\eta_{UROP}\right\} \leq 1
\end{equation}
\begin{equation}\label{69}
1- \frac{k E\left\{N-T_0\right\}}{V_{tot}(N)} \leq E\left\{\eta_{UROP}\right\} \leq 1
\end{equation}
We denote by $\tau_{ar,i}$ and $\tau_{dep,i}$, elapsed time between two consecutive selection of same sensor $s_i$ and elapsed time between two consecutive departure of same sensor $s_i$. For long problem horizons, $E\left\{\tau_{ar,i}\right\}$=$E\left\{\tau_{dep,i}\right\}, \forall i$. By Lemma 3, $L\subset(F_1\cup F_2)$. By Lemma 2, none of nodes $s_i\in S_m-S_k$ can be selected (started to schedule) more than twice by the FC in the interval $[T_0,N]$; therefore, $E\left\{N-T_0\right\}<2E\left\{\tau_{ar}\right\}$. None of the nodes $s_i\in S_k$ can be left (stopped to schedule) more than once by the FC in interval $\left[\zeta_i^{(f)},N\right]$; therefore, $E\left\{N-\zeta_i^{(f)}\right\}<2E\left\{\tau_{dep}\right\}$. Hence, (\ref{69}) is converted into (\ref{70}):
\begin{equation}\label{70}
1- \frac{2kE\left\{\tau_{ar}\right\}}{V_{tot}(N)} \leq E\left\{\eta_{UROP}\right\} \leq 1
\end{equation}
Let denote by $D$ and $K$, density during problem horizon $N$ and the number of orthogonal channels of the FC. By definition of $D$, $V_{tot}(N)=DNk$.
\begin{equation}\label{71}
D=\frac{kE\left\{\tau_{ar}\right\}-m}{kE\left\{\tau_{ar}\right\}}
\end{equation}
\begin{equation}\label{72}
E\left\{\tau_{ar}\right\}=\frac{m}{k(1-D)}
\end{equation}
\begin{equation}\label{73}
1-\frac{2k\frac{m}{k(1-D)}}{DNk}<E\left\{\eta_{UROP}\right\}\leq 1
\end{equation}
\begin{equation}\label{74}
1-\frac{2m}{\left(1-D\right)DNk}<E\left\{\eta_{UROP}\right\}\leq 1
\end{equation}
\textbf{\textit{Note:}} Since $D=0$ means no harvested energy in the whole network, it is trivial case and not considered in our calculations. $D=1$ means that there is no idle TS if FC apply the 100\% efficient policy ($\pi^{fe}$). However, UROP benefits from idle TSs to schedule the sensors. From Theorem 1, no $\pi^{fe}$ exists for $D>1$. Therefore, we investigate $0<D<1$ in this paper.

\section{Extension to the Infinite-Horizon Case}
As  in (\ref{23}) and (\ref{29}), efficiency of RR-based policies (also MP in \cite{18} and \cite{20}) depend on not only sensor densities $D$ and $D^{(T)}$ but also partial sensor densities $D_i$ and $D_i^{(T)}$ and cannot improve as problem horizon goes to infinity. Also, it is proved that batteries of nodes for which $D_i>1$ and $D_i^{(T)}>1$ will overflow over infinite horizon. However, efficiency of UROP in finite horizon case improves as the problem horizon increase and goes to infinity. By Theorem 5 and the relation $V_{tot}^{opt}(N)=DNk$, efficiency of UROP is, for $0<D<1$, 
\begin{equation}\label{75}
\lim_{N\rightarrow \infty} \left(1-\frac{2m}{\left(1-D\right)DNk}\right)<\lim_{N\rightarrow \infty} E\left\{\eta_{UROP}\right\}\leq 1
\end{equation}
Hence, $\lim_{N\rightarrow \infty}E\left\{\eta_{UROP}\right\}=1$, which shows that UROP is \textit{asymptotically optimal} in the infinite horizon for general EH processes.

\section{Numerical Results}

In this section, efficiency achieved by RR and UROP policies are compared for independent (Poisson) and correlated (Markovian) EH processes under high density (D=0.975) and low density (D=0.2) EH processes  first. RR and UROP are then compared under a fairness criterion for independent (Poisson) and correlated (Markovian) EH processes under high density (D=0.975). Finally, computational complexities of RR, UROP and UP (the omniscient policy proposed in section V) are compared. We focus on the region $D^{(T)}\leq 1$ so $\eta_{opt}=\eta_{fe}=1$. 

To begin with, we compare efficiencies of these policies under both infinite and finite battery assumption for four cases. To create a realistic scenario, we take $m=100$, $k=10$, $N=2000$ for both policies. We also investigate the efficiency of UROP by taking $m=103$ and $k=10$. Note that we compare efficiency of UROP with an arbitrary RR since $\eta_{RR}\cong\eta_{MP}$ for long problem horizons (Theorem 3). We investigate the efficiencies of both policies under a nonuniform EH process (Both achieve nearly 100\% efficiency for uniform EH processes). Nonuniform, high density traffic is formed by taking $D_i=3$ for 25 of the nodes and $D_i=0.3$ for the remaining ones. Moreover, low density, nonuniform traffic is formed by taking $D_i=2.1$ for 5 nodes and $D_i=0.1$ for the remaining nodes. Independent EH processes are modelled as Poisson. Markov EH process are modelled by a state space $M_i=\left\{0, 1, 2\right\}$, $\forall s_i$ and a $3\times 3$ transition matrix $P$ such that $P_{ii}=0.9$ $\forall i$ and $P_{ij}=0.05$ for $i\neq j$. The harvested energy for node $s_i$ in TS $t$, $E_i^h(t)$, is determined by $M_i$ such that $E_i^h(t)=D_i\times M_i(t)$ (Note that each transmission requires unit energy.).

In Figure 5 (Low density, independent EH process), UROP has nearly 100\% efficiency whereas RR has approximately 80\% efficiency. In Figure 6 (High density, independent EH process), UROP continues to attain nearly 100\% efficiency whereas the efficiency of RR has dropped below 50\%. This is an expected result since Theorem 2 states that as the number of nodes s.t. $D_i>1$ increases, efficiency of RR decreases. By (\ref{20}), efficiency of RR is expected to be $\eta_{RR}=48.7\%$ and $\eta_{RR}=72.5\%$ for the low and high density EH process, respectively.

In Figure 7 (Low density, Markov EH process), UROP has nearly 100\% efficiency whereas RR has nearly 70\% efficiency. In Figure 8 (High density, Markov EH process), UROP has nearly 100\% efficiency whereas RR has nearly 50\% efficiency. When the EH process has memory, we observe similar results, except that the performance of RR drops further. The efficiency of UROP is more robust to memory in harvest process, as compared to RR (Note that $P_{ii}=0.9$, $\forall i$).

Considering all four figures, we wish to make three additional remarks. First, the efficiency of UROP converges to 100\% $N \rightarrow \infty$, as shown in Section VII (UROP is asymptotically optimal). Secondly, efficiency of UROP with a reasonable-sized finite battery $B_i$=50 is almost same as that with infinite battery. Finally, UROP can achieve nearly 100\% throughput both for $m/k\in Z$ and $m/k\notin Z$ cases, while RR needs $m/k\in Z$ assumption for optimality. We conclude that UROP is more adaptive and efficient than RR (and MP proposed in \cite{18, 20} by Theorem 3).

In addition to throughput, the performances of RR and UROP are also compared in terms of fairness, which is often an important issue for scheduling policies. We apply Jain's Fairness index \cite{29},
$f(x)=\frac{[\sum_{i=1}^m x_i(t)]^2}{m \sum_{i=1}^m x_i^2(t)}$
where $x_i(t)$  is the $i^{th}$ user allocation up to TS $t$. Adopting the \emph{proportionate progress} (P-fairness) criterion in \cite{30}, we scale the allocation $x_i(t)=\frac{V_i(t)}{V_i^{opt}(t)}$ over users.

RR is usually known as a fair policy since it schedules users periodically. RR is 100\% fair for uniform EH processes. However, RR may not be very fair for nonuniform EH processes. In fact, from \ref{23} and \ref{29} the efficiency of RR is expected to be $FI_{RR}=89.3\%$ for high density $D=0.975$, nonuniform arrivals. On the other hand, UROP schedules the users proportionally to their loads as well as respecting same or periodically. Consequently, UROP can achieve 100\% fairness for general case of EH process. This is evident on Figure 9 and Figure 10. It is also observed that UROP is nearly 100\% fair also for $m/k$ noninteger case. 

In addition to throughput and fairness, the policies are compared in terms of computational complexity. RR has complexity $O(1)$. Besides achieving almost 100\% throughput and 100\% fairness for various EH processes, UROP has low-complexity as well. In each TS, UROP checks the $k$ nodes which are scheduled in previous TS thus it makes only $k$ computations in each TS. Therefore, computational complexity of UROP is $O(kN)$. It is also interesting to compare UROP with UP in terms of computational complexity. UP is an optimal omniscient policy. In each TS, UP checks the $k$ nodes scheduled in previous TS and looks for replacement nodes if some of the $k$ nodes cannot transmit data in that TS. Number of computation which UP makes in each TS is between $k$ and $m$. Hence, UP has a computational complexity between $O(kN)$ and $O(mN)$.  The results show that to achieve 100\% throughput, UP may have complexity $O(mN)$ whereas UP may have complexity $O(kN)$. This implies that UP may have $\frac{m}{k}$ times more computation than UROP to achieve the same throughput performance. 
In other words, UROP achieves the same performance as UP with up to $\frac{m}{k}$ times lower complexity.

\section{Conclusion}

This paper investigated a scheduling problem for a single-hop WSN where a fusion center(FC) schedules a set of EH nodes to receive data from them. FC does not know the instantaneous battery states of nodes. Batteries get recharged according to random Energy Harvesting processes, whose statistics are not available to the FC, and there is no leakage from the batteries. Under an infinite battery capacity assumption, we exhibit a near-optimal online scheduling policy for a broad set of EH processes (Markovian, independent, uniform, nonuniform, etc.) 

The scheduling problem is set up as an expected undiscounted reward maximization problem. It is shown that Round Robin (RR) based policies are generally suboptimal(do not guarantee 100\% throughput) for nonuniform EH processes. It is also shown that policies proposed in previous literature (namely, myopic policies in \cite{18} and \cite{20} have almost equal efficiency as any other RR policy with quantum=1. 

Next, a low-complexity scheduling policy, UROP, is proposed. It is shown that UROP is asymptotically optimal regardless of traffic, in the infinite horizon. Even in the finite horizon, UROP achieves nearly 100\% throughput without requiring feedback about battery states of nodes. As this is a type of restless multi-armed bandit problem, the simple self-adapting scheduling technique of UROP could find potential applications in problems other than communication networks, whenever the performance measure is average reward and the queues store a flexible (time insensitive) resource such as energy.


%





\ifCLASSOPTIONcaptionsoff
  \newpage
\fi



%

\begin{figure}
	\centering	\includegraphics[width=0.82\textwidth]{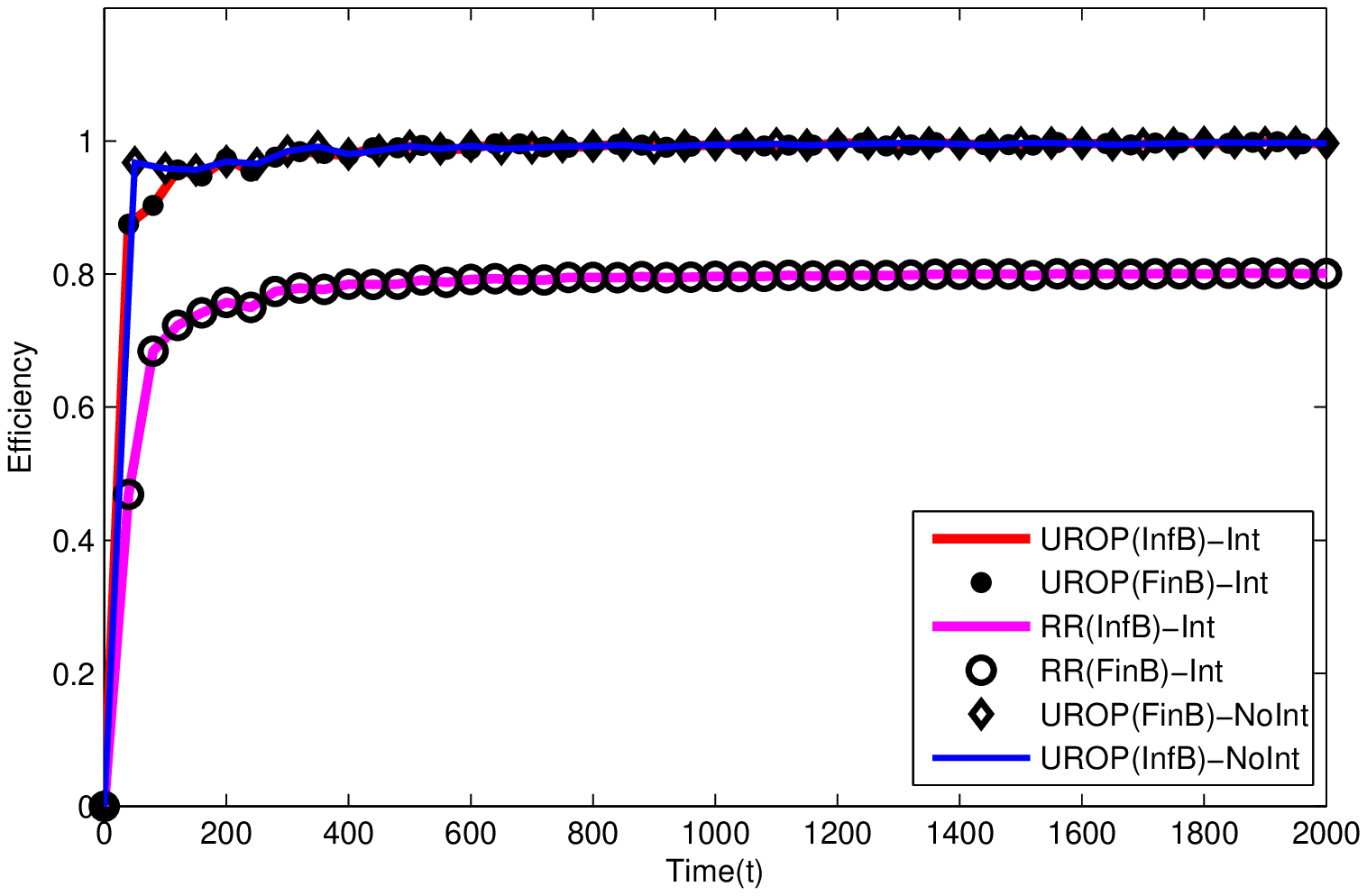}
	\caption{Efficiencies (ratio of total throughput by a policy to total throughput by optimal policy) of UROP, RR under infinite and finite battery $B_i$=50 assumptions for independent low density energy arrivals ($D=0.2$) such that $m/k\in Z$. Efficiency of UROP is also shown for $m/k$ taking a noninteger value.}
	\label{Figure 5}
\end{figure}
\begin{figure}
	\centering	\includegraphics[width=0.82\textwidth]{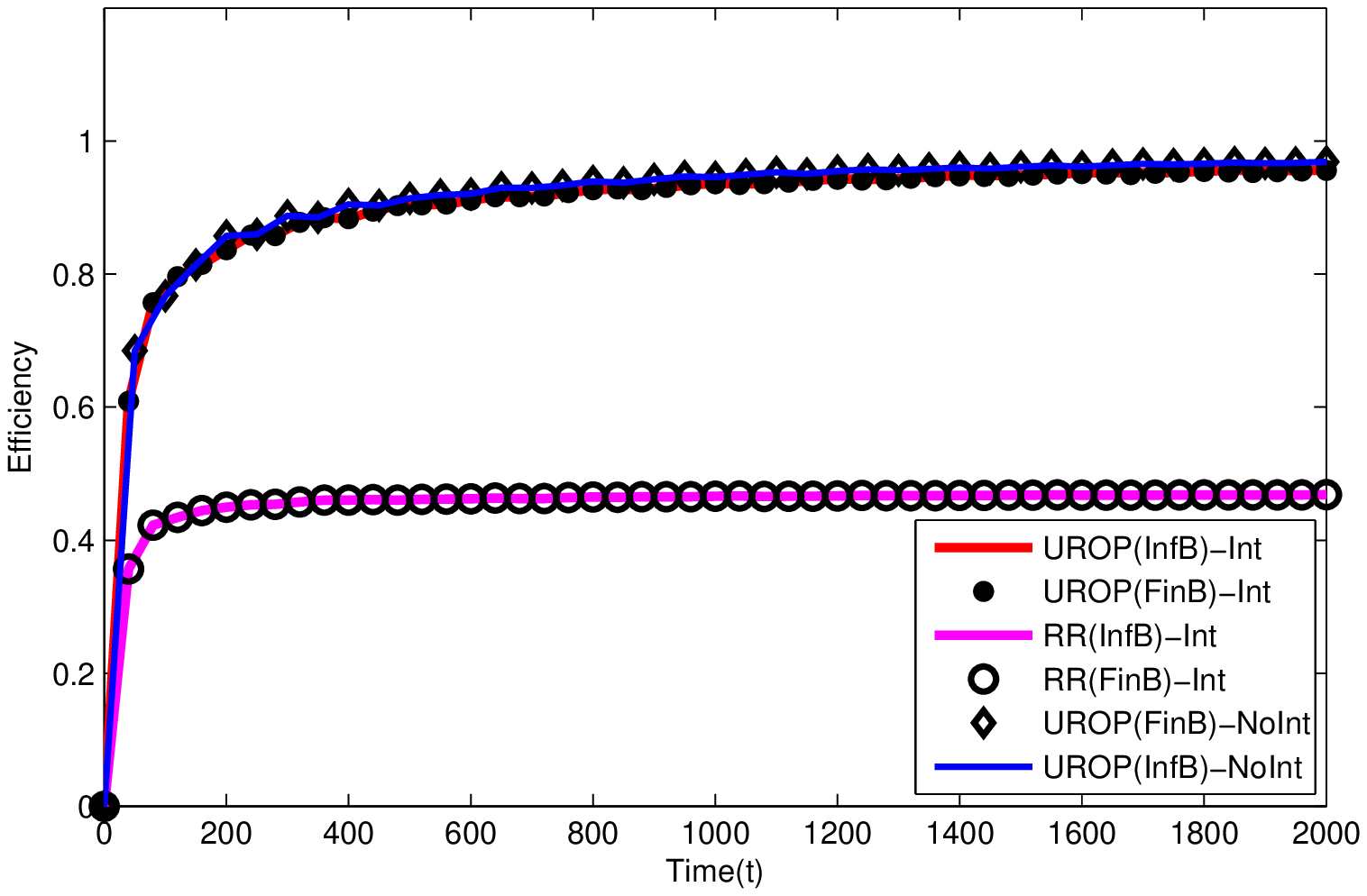}
	\caption{Efficiencies (ratio of total throughput by a policy to total throughput by optimal policy) of UROP, RR under infinite and finite battery $B_i$=50 assumptions for independent high density energy arrivals ($D=0.975$) such that $m/k\in Z$. Efficiency of UROP is also shown for $m/k$ taking a noninteger value.}
	\label{Figure 6}
\end{figure}
\begin{figure}
	\centering	\includegraphics[width=0.82\textwidth]{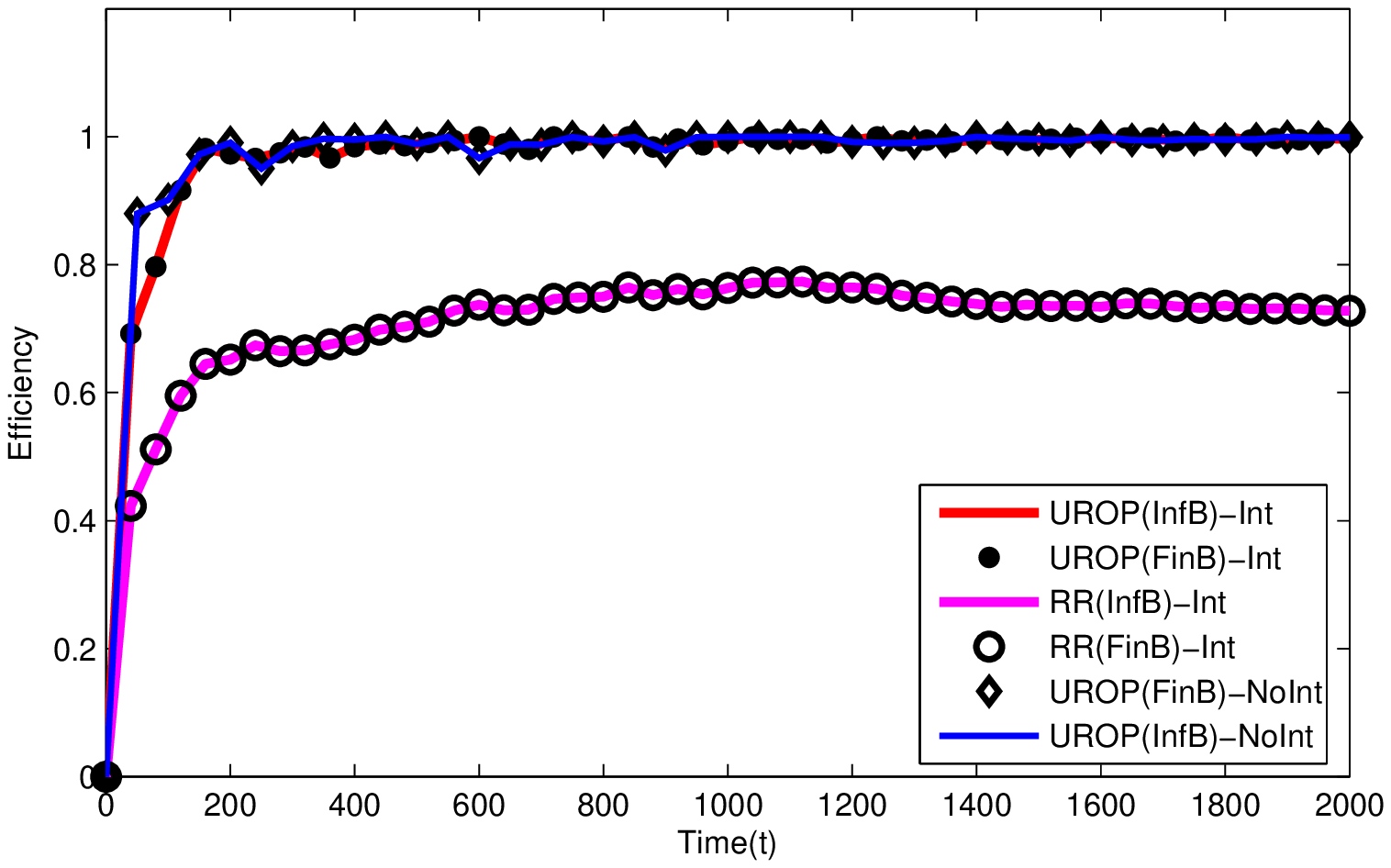}
	\caption{Efficiencies (ratio of total throughput by a policy to total throughput by optimal policy) of UROP, RR under infinite and finite battery $B_i$=50 assumptions for Markov low density energy arrivals ($D=0.2$) such that $m/k\in Z$. Efficiency of UROP is also shown for $m/k$ taking a noninteger value.}
	\label{Figure 7}
\end{figure}
\begin{figure}
	\centering	\includegraphics[width=0.82\textwidth]{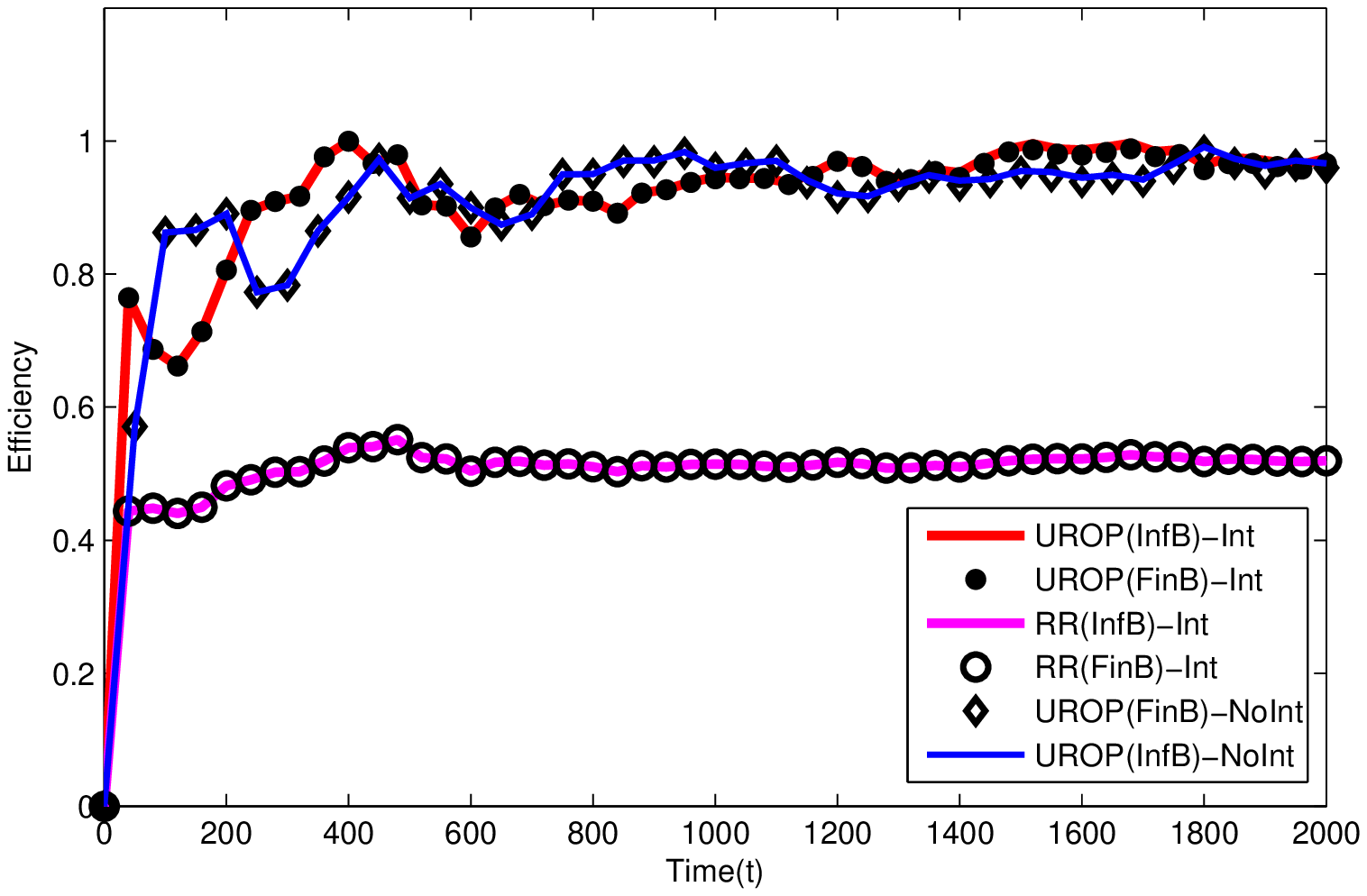}
	\caption{Efficiencies (ratio of total throughput by a policy to total throughput by optimal policy) of UROP, RR under infinite and finite battery $B_i$=50 assumptions for Markov low density energy arrivals ($D=0.975$) such that $m/k\in Z$. Efficiency of UROP is also shown for $m/k$ taking a noninteger value.}
	\label{Figure 8}
\end{figure}
\begin{figure}
	\centering	\includegraphics[width=0.82\textwidth]{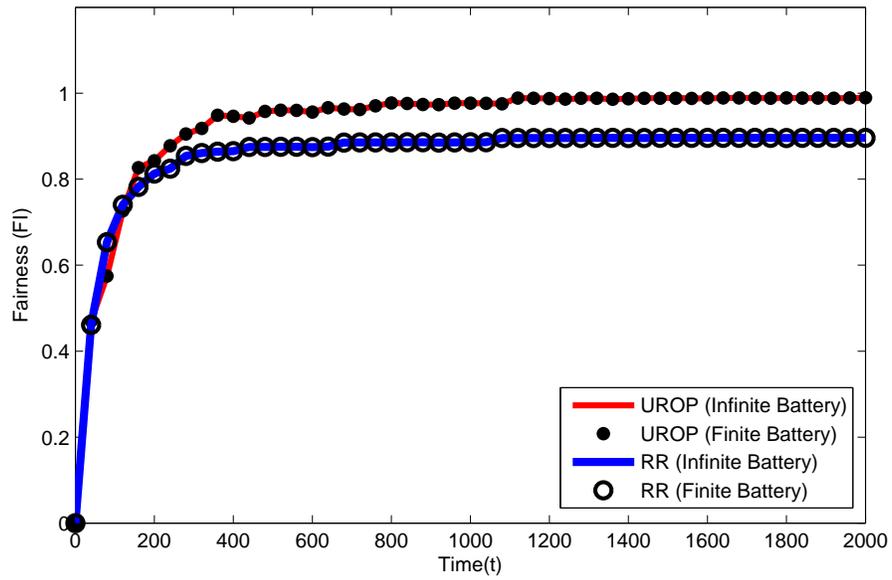}
	\caption{Fairness of UROP, RR under infinite and finite battery $B_i$=50 assumptions for high density $D=0.975$ and independent EH process by $m/k$ integer assumption.}
	\label{Figure 9}
\end{figure}
\begin{figure}
	\centering	\includegraphics[width=0.82\textwidth]{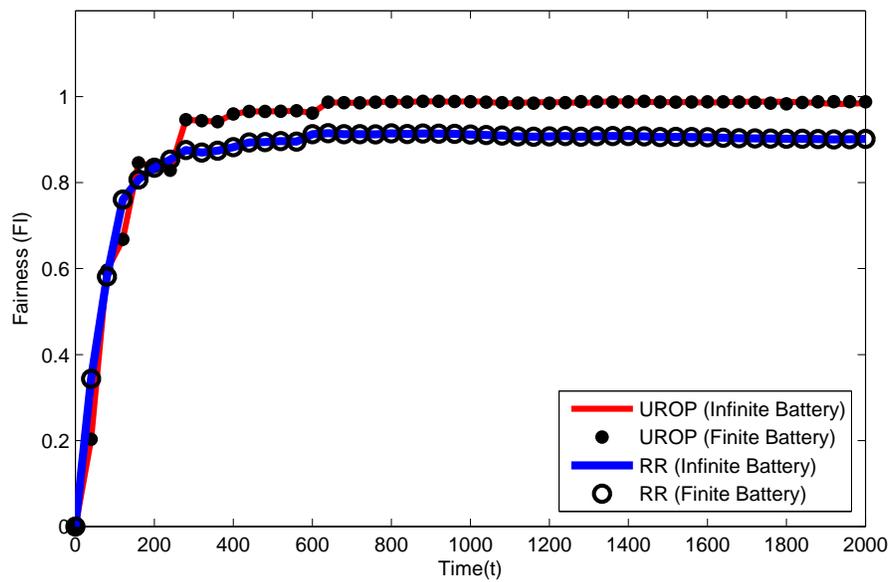}
	\caption{Fairness of UROP, RR under infinite and finite battery $B_i$=50 assumptions for high density $D=0.975$ and Markov EH process by $m/k$ integer assumption.}
	\label{Figure 10}
\end{figure}

%








\end{document}